\documentclass[12pt,preprint]{elsarticle} 

\usepackage[utf8]{inputenc}
\usepackage{amsmath, amssymb, amsfonts}
\usepackage{geometry}
\usepackage{graphicx}
\usepackage{indentfirst}
\usepackage{caption}
\usepackage{subcaption}
\usepackage{xcolor}
\usepackage{placeins}
\begin{document}

\begin{frontmatter}

\title{Harmonic Balance Unified Gas-Kinetic Scheme for Multiscale Periodic Flows}

\author[a]{Yuze ZHU}
\author[a]{Hangkong WU}
\author[a]{Yufeng WEI}
\author[a,b,c]{Kun XU\corref{cor1}}
\cortext[cor1]{Corresponding author. Email: makxu@ust.hk}

\affiliation[a]{
   organization = {Department of Mathematics, Hong Kong University of Science and Technology},
   city         = {Clear Water Bay, Kowloon, Hong Kong},
   country      = {China},
}

\affiliation[b]{
   organization = {Department of Mechanical and Aerospace Engineering, Hong Kong University of Science and Technology},
   city         = {Clear Water Bay, Kowloon, Hong Kong},
   country      = {China},
}

\affiliation[c]{
   organization = {Shenzhen Research Institute, Hong Kong University of Science and Technology},
   city         = {Shenzhen},
   country      = {China},
}

\begin{abstract}

A time-domain harmonic balance unified gas-kinetic scheme (HB-UGKS) is developed for simulating periodic non-equilibrium flows across all Knudsen regimes. By applying a time-spectral operator, the unsteady periodic problem is reformulated into a block-coupled, quasi-steady system. To preserve the intrinsic multiscale transport–collision flux coupling of the UGKS, a source-separated formulation incorporates the harmonic balance coupling strictly at the cell-residual level. This allows the periodic limit cycle to be resolved directly via pseudo-time marching with local time-stepping, advancing all temporal collocation points simultaneously and completely bypassing physical startup transients. The framework is validated against three complementary benchmarks: two small-amplitude shear-driven oscillatory flows and a thermally driven cavity under finite-amplitude excitation. The scheme accurately captures intricate non-equilibrium kinetic phenomena, including aspect-ratio anti-resonance scaling, dynamic shear traction, nonlinear waveform distortions, and acoustic streaming. Across all cases, the HB-UGKS preserves the fidelity of standard time-accurate simulations, while achieving order-of-magnitude speedups in high-frequency regimes where long transient timescales dominate.
\end{abstract}

\end{frontmatter}

\section{Introduction}

Periodic gas flows are critical to a wide range of micro-mechanical applications, ranging from thermally-driven Knudsen compressors to mechanically-oscillating resonators. In these micro- and nano-electromechanical systems (MEMS/NEMS), characteristic length scales frequently approach the molecular mean free path, rendering rarefaction effects unavoidable\cite{LIGHTHILL1978391,annurev:/content/journals/10.1146/annurev.fluid.30.1.579,cercignani2000rarefied,sone2007molecular}. Beyond spatial confinement, the influence of high-frequency excitations further drives the gas far from local thermodynamic equilibrium, manifesting in significant multiscale characteristics. Since the precise characterization of these multiscale effects is vital for predicting device performance and ensuring their operational reliability, developing a numerical framework that remains both accurate and computationally efficient is of paramount importance\cite{Beskok1999,10.1063/1.1874193}.

To resolve these multiscale non-equilibrium flows, various numerical frameworks have been developed based on the Boltzmann equation and its kinetic models, such as the Bhatnagar–Gross–Krook (BGK) model \cite{bhatnagar1954model}. These methods are generally categorized into stochastic particle methods and deterministic approaches. Among the former, the direct simulation Monte Carlo (DSMC) method \cite{bird1963approach,bird1970direct, bird1994molecular} is the most representative; however, its application to micro-scale periodic flows is often plagued by significant statistical noise, particularly in low-speed regimes \cite{10.1063/1.2393436}. To mitigate these issues, variants such as the information preservation (IP) method \cite{FAN2001393} and low-variance DSMC \cite{HOMOLLE20072341} have been proposed. In addition to stochastic techniques, deterministic approaches have been extensively developed for rarefied gas dynamics. While moment methods \cite{doi:10.1137/100785466} offer high computational efficiency, they frequently suffer from stability and closure challenges in the transition regime. In contrast, discrete velocity methods (DVM) \cite{chu1965kinetic,Broadwell_1964,MIEUSSENS2000429} provide a more robust alternative, with significant advances including fast spectral methods \cite{WU2015602,WU2019108924} for efficient evaluation of collision integrals and the general synthetic iterative scheme (GSIS) \cite{ZHU2021110091} for accelerated convergence. Despite these developments, most conventional kinetic schemes—whether stochastic or deterministic—still rely on operator splitting to decouple transport and collision. This decoupling strictly constrains the grid size and time step to the molecular mean free path and collision time, respectively. To address these issues, Xu et al. proposed the unified gas-kinetic scheme (UGKS) \cite{xu2010unified,huang2012unified}, which couples molecular transport and collision within the flux evaluation. By resolving the flow physics across all flow regimes—from the continuum limit to the free-molecular limit—the UGKS and its variants, such as the discrete unified gas-kinetic scheme (DUGKS) \cite{guo2021progress}, have demonstrated remarkable efficacy in modeling diverse multiscale flows\cite{Liu2017Unified,XIAO2017475,SUN2015222,LIU2019264}.

Despite its success in multiscale modeling, the UGKS—much like other conventional numerical frameworks—typically resolves unsteady flows through a time-marching strategy. However, the solution accuracy and computational efficiency of such simulations are significantly influenced by the specification of the physical time step: while an excessively large step compromises the resolution of transient flow features, an overly small step leads to an escalation in computational expenditure and memory requirements. Particularly for periodic unsteady flows, the system usually requires marching through numerous cycles to eliminate initial transients before reaching a periodic steady state\cite{10.1115/1.4025203, blazek2015computational}. This process is exceptionally resource-intensive for kinetic schemes like the UGKS, which involve high-dimensional discrete velocity spaces. Although various implicit and convergence acceleration techniques have achieved significant progress over the past decade\cite{10.1063/1.4994020,ZHU2019190,tan2020time,xu2022ugks}, the trade-off between temporal resolution and computational efficiency remains a substantial challenge for the rapid design iteration of complex micro-systems when handling high-frequency periodic excitations.

To circumvent the inherent limitations of conventional time-marching strategies, an effective approach is to leverage Fourier theory to transform the original unsteady governing equations into a time-independent, quasi-steady system. Following this concept, frequency-domain linearized methods have been widely applied across aerodynamics, aeroelasticity, and gas kinetics\cite{hall1993linearized,clark2000time,wu2014oscillatory}, demonstrating exceptional efficiency and precision in resolving linear flow fields. Specifically within the realm of gas kinetics, Wu et al.~\cite{wu2014oscillatory} successfully solved for the complex Fourier coefficients of the perturbed distribution function. However, the fundamental reliance on a small-perturbation assumption to linearize the collision operator strictly confines this approach to linear oscillatory flows, rendering it incapable of capturing complex nonlinear flow behaviors induced by finite-amplitude excitations. Furthermore, the complex mathematical derivation of specific linearized operators severely limits its general extensibility.

In contrast, the time-domain harmonic balance (HB) method provides an effective alternative that bridges the gap between high efficiency and nonlinear capability\cite{hall2002computation,hall2013harmonic}. By dividing the period into several temporal collocation points and coupling their solutions through a time spectral operator, the HB method achieves a similar Fourier-based transformation into a quasi-steady system. However, because the HB method directly evaluates the original nonlinear governing equations at each temporal collocation point, it naturally captures finite-amplitude nonlinearities without invoking any small-perturbation assumptions. This transformation completely bypasses the period-by-period transient marching, demonstrating a remarkable capacity for accelerated convergence \cite{10.1115/1.4068947}. Furthermore, by avoiding the intrusive mathematical reformulation associated with linearized operators, the HB method exhibits inherent parallel scalability and can be straightforwardly integrated into existing steady or unsteady solvers \cite{10.1115/1.4066586,huang2019time,huang2018implicit}. Although the HB method has been extensively utilized within the Navier--Stokes (NS) framework for continuum flows, its theoretical extension and application to multiscale, all-regime periodic flows remain largely unexplored. Consequently, combining the all-regime physical fidelity of the UGKS with the high efficiency and nonlinear capability of the HB method holds great significance for the performance evaluation of precision MEMS devices and the exploration of microscopic non-equilibrium flow mechanisms.

In this work, we develop a time-domain harmonic balance unified gas-kinetic scheme (HB-UGKS) for periodic gas flows across all Knudsen regimes. Three representative benchmarks are chosen to systematically evaluate the numerical accuracy, physical fidelity, and computational efficiency of the proposed framework. First, a small-amplitude shear-driven cavity flow is examined across various frequencies, Knudsen numbers, and aspect ratios, establishing the baseline linear accuracy and capturing acoustic anti-resonance phenomena against spectral solutions of the linearized Boltzmann equation \cite{wu2014oscillatory}. Second, to further examine the multiscale shear transport and phase lag across diverse Knudsen regimes, an oscillatory shear-driven flow in an eccentric annulus is resolved and benchmarked against reference solutions obtained via the frequency-domain general synthetic iterative scheme (GSIS) \cite{li2026frequency}. Third, a large-amplitude thermally driven cavity is simulated to assess the solver's performance in capturing finite-amplitude non-equilibrium nonlinearities, where the necessity of higher-order harmonics, temporal waveform distortions, and dynamic Knudsen layers are verified against both time-marching and reference data \cite{Yang2025Hybrid}. Furthermore, to substantiate the computational advantages enabled by the HB formulation, each benchmark incorporates systematic paired comparisons against the conventional time-domain UGKS (TD-UGKS) under identical discretization conditions, highlighting the speedup achieved in capturing periodic steady states.

The remainder of this paper is organized as follows. Section 2 presents the kinetic model and the HB-UGKS formulation. Section 3 evaluates the solver using three benchmarks: a linear shear-driven cavity, an oscillatory eccentric annulus, and a finite-amplitude thermally driven cavity, with accuracy, flow physics, and computational efficiency analyzed in each case. Section 4 summarizes the conclusions.

\section{Methodology}

\subsection{Kinetic model equation}

In the kinetic theory of gases, the evolution of a gas flow is fundamentally governed by the Boltzmann equation. To alleviate the prohibitive computational cost associated with evaluating the full collision integral, various kinetic relaxation models have been developed. In the present study, the Shakhov-modified BGK model \cite{shakhov1968generalization} is adopted to accurately resolve the heat flux with an arbitrary Prandtl number. The standard BGK model \cite{bhatnagar1954model}, which serves as the baseline for this relaxation process, can be expressed as:

\begin{equation}
\frac{\partial f}{\partial t} + \mathbf{u}\cdot \nabla_{\mathbf{x}} f = \frac{g - f}{\tau},
\label{eq:kinetic_relax}
\end{equation}
where $f(\mathbf{x},\mathbf{u},\boldsymbol{\xi},t)$ is the particle velocity distribution function, and $\mathbf{x}$, $\mathbf{u}$, $\boldsymbol{\xi}$, and $t$ represent the physical space coordinate, microscopic particle translational velocity, velocity variables for internal degrees of freedom, and time, respectively.  The local relaxation time $\tau$ is related to the macroscopic viscosity $\mu$ and pressure $p$ via:
\begin{equation}
    \tau = \mu/p.
\end{equation}

The local equilibrium state $g$ is given by the Maxwellian distribution:
\begin{equation}
g = \rho \left(\frac{\lambda}{\pi}\right)^{\frac{D+K}{2}} e^{ -\lambda \left( |\mathbf{u}-\mathbf{U}|^2 + \boldsymbol{\xi}^2 \right)},
\label{eq:maxwellian}
\end{equation}
where $\rho$ is the macroscopic density, $\mathbf{U}$ is the macroscopic velocity vector, and $\lambda = 1/(2RT)$, with $R$ and $T$ being the specific gas constant and the absolute temperature, respectively. The parameter $K$ denotes the number of internal degrees of freedom. For a monatomic gas (e.g., Argon with specific heat ratio $\gamma = 5/3$) simulated in a two-dimensional physical space ($D=2$), the out-of-plane translational velocity component is effectively treated as an internal degree of freedom, yielding $K=1$.

A well-known deficiency of the BGK model is its unity Prandtl number ($\Pr=1$). To accurately resolve the thermal boundary layer and recover the correct Prandtl number ($\Pr \approx 2/3$ for monatomic gases), the Shakhov model \cite{shakhov1968generalization} modifies the Maxwellian equilibrium into $g^{+}$:
\begin{equation}
g^{+} = g\left[ 1+(1-\Pr)\frac{\mathbf{c}\cdot\mathbf{q}}{5pRT} \left(\frac{|\mathbf{c}|^2+\boldsymbol{\xi}^2}{RT}-5\right) \right],
\label{eq:shakhov}
\end{equation}
where $\mathbf{c} = \mathbf{u} - \mathbf{U}$ is the peculiar velocity vector, $p = \rho RT$ is the static pressure, and $\mathbf{q}$ is the heat flux vector. In the remainder of this paper, $g^{+}$ and $g$ refer to the Shakhov-corrected equilibrium and the baseline Maxwellian, respectively.

The macroscopic conservative variables $\mathbf{W} = (\rho, \rho \mathbf{U}, \rho E)^T$ are obtained by taking the moments of the distribution function $f$ over the velocity space:
\begin{equation}
\mathbf{W} = \int \boldsymbol{\psi} f d\Xi,
\label{eq:moment}
\end{equation}
where $\boldsymbol{\psi} = \left( 1, \mathbf{u}, \frac{1}{2}(|\mathbf{u}|^2+\boldsymbol{\xi}^2) \right)^T$ is the vector of collision invariants, and $d\Xi = d\mathbf{u}\,d\boldsymbol{\xi}$ is the velocity-space volume element. Correspondingly, the macroscopic heat flux $\mathbf{q}$ is evaluated as follows:
\begin{equation}
\mathbf{q} = \frac{1}{2}\int \mathbf{c}\left(|\mathbf{c}|^2+\boldsymbol{\xi}^2\right) f d\Xi.
\label{eq:heatflux}
\end{equation}

In the discrete-velocity framework adopted herein, the continuous velocity space is discretized into a finite set of discrete velocity points $\{\mathbf{u}_\alpha\}$ with corresponding quadrature weights $\{w_\alpha\}$. To eliminate the computational burden of resolving the multidimensional internal velocity space, the internal velocity variable $\boldsymbol{\xi}$ is analytically integrated out by introducing two reduced distribution functions:
\begin{equation}
h(\mathbf{x},\mathbf{u},t) = \int f d\boldsymbol{\xi},
\label{eq:reduced_h}
\end{equation}
\begin{equation}
b(\mathbf{x},\mathbf{u},t) = \int \boldsymbol{\xi}^2 f d\boldsymbol{\xi}.
\label{eq:reduced_b}
\end{equation}

With the reduced distributions $(h,b)$, the
conservative variables and heat flux can be evaluated without explicitly discretizing the
internal velocity variable:
\begin{equation}
\mathbf{W}
=
\sum_{\alpha} w_\alpha
\begin{bmatrix}
h_\alpha \\
\mathbf{u}_\alpha h_\alpha \\
\dfrac{1}{2}
\left(
|\mathbf{u}_\alpha|^2 h_\alpha+b_\alpha
\right)
\end{bmatrix},
\label{eq:reduced_moments}
\end{equation}
and
\begin{equation}
\mathbf{q}
=
\frac{1}{2}
\sum_{\alpha} w_\alpha
\mathbf{c}_\alpha
\left[
|\mathbf{c}_\alpha|^2 h_\alpha+b_\alpha
\right],
\label{eq:reduced_heat_flux}
\end{equation}
where
$\mathbf{c}_\alpha=\mathbf{u}_\alpha-\mathbf{U}$.

To incorporate the Shakhov correction into the relaxation model, the reduced target equilibrium states are defined as:
\begin{equation}
h^+ = H + H^+,
\end{equation}
\begin{equation}
b^+ = B + B^+,
\label{eq:reduced_equilibrium}
\end{equation}
where $H = \int g\, d\boldsymbol{\xi}$ and $B = \int \boldsymbol{\xi}^2 g\, d\boldsymbol{\xi}$ denote the reduced Maxwellian distributions. The correction terms, $H^+ = \int (g^+ - g)\, d\boldsymbol{\xi}$ and $B^+ = \int \boldsymbol{\xi}^2 (g^+ - g)\, d\boldsymbol{\xi}$, account for the non-unity Prandtl number and are expressed as functions of the macroscopic heat flux. Consequently, the high-dimensional velocity-space integrals are effectively reduced to $D$-dimensional numerical summations over the discrete velocity set $\{\mathbf{u}_\alpha\}$. The evolution of the discrete-kinetic system is thus fully determined by the reduced distributions $h$ and $b$.

\subsection{Time-domain harmonic balance unified gas-kinetic scheme}

For periodic flows with an angular frequency $\omega$ and period $T = 2\pi/\omega$, the asymptotic periodic state is the primary focus rather than the initial start-up transients. In contrast to direct time-marching solutions, the time-domain HB method converts the unsteady governing equations into a coupled quasi-steady system by representing the period over a finite set of temporal collocation points for both the macroscopic and microscopic update equations. This approach transforms temporal derivatives into a global spectral source operator while preserving the original multiscale characteristics of the UGKS. Notably, the time spectral operator is incorporated into both the macroscopic and microscopic equations to update the conservative variables and distribution functions. Therefore, within the HB-UGKS framework, the solution vector—comprising macroscopic variables $\mathbf{W}$ and reduced distributions $(h, b)$—is solved concurrently across all temporal collocation points. Following the Nyquist–Shannon sampling theorem, any time-dependent variable is represented by a truncated complex Fourier series with $N_\text{H}$ harmonics. Specifically, the macroscopic conservative variables and microscopic distribution functions are expressed as:
\begin{subequations}
\label{eq:fourier_series}
\begin{equation}
\mathbf{W}(\mathbf{x}, t) \approx \sum_{k=-N_\text{H}}^{N_\text{H}} \widehat{\mathbf{W}}_k(\mathbf{x})\, e^{i k \omega t},
\label{eq:fourier_W}
\end{equation}
\begin{equation}
h(\mathbf{x}, \mathbf{u}, t) \approx \sum_{k=-N_\text{H}}^{N_\text{H}} \widehat{h}_k(\mathbf{x}, \mathbf{u})\, e^{i k \omega t},
\label{eq:fourier_h}
\end{equation}
\begin{equation}
b(\mathbf{x}, \mathbf{u}, t) \approx \sum_{k=-N_\text{H}}^{N_\text{H}} \widehat{b}_k(\mathbf{x}, \mathbf{u})\, e^{i k \omega t},
\label{eq:fourier_b}
\end{equation}
\end{subequations}
where $i$ denotes the imaginary unit. The terms $\widehat{\mathbf{W}}_k$, $\widehat{h}_k$, and $\widehat{b}_k$ represent the complex Fourier coefficients of the $k$-th harmonic for the macroscopic conservative variables and the reduced distribution functions, respectively. Since the physical flow variables are real-valued, their Fourier coefficients satisfy the conjugate symmetry property, i.e., $\widehat{q}_{-k} = \widehat{q}_k^*$ for any variable $q \in \{\mathbf{W}, h, b\}$, where the superscript $*$ denotes the complex conjugate.

To avoid the complex nonlinear convolution operations associated with the kinetic collision term in a fully frequency-domain approach, a time-domain collocation strategy is adopted. The temporal period $T$ is discretized into $N_\text{T} = 2N_\text{H} + 1$ equidistant temporal collocation points:
\begin{equation}
t_n = \frac{n T}{N_\text{T}}, \qquad n = 0, 1, \dots, 2N_\text{H}.
\label{eq:collocation_points}
\end{equation}

By assembling the flow variables at these collocation points into time-domain vectors, e.g., $\mathbf{W}^* = [\mathbf{W}|_{t_0}, \dots, \mathbf{W}|_{t_{2N_\text{H}}}]^T$, the discrete Fourier transform (DFT) provides a direct mapping between the solution values at the discrete temporal collocation points and their corresponding frequency-domain Fourier coefficients:
\begin{subequations}
\label{eq:dft_mapping}
\begin{equation}
\widehat{\mathbf{W}} = \mathbf{D}\,  \mathbf{W^*},
\label{eq:dft_W}
\end{equation}
\begin{equation}
\widehat{\mathbf{h}} = \mathbf{D}\,  \mathbf{h^*},
\label{eq:dft_h}
\end{equation}
\begin{equation}
\widehat{\mathbf{b}} = \mathbf{D}\,  \mathbf{b^*}.
\label{eq:dft_b}
\end{equation}
\end{subequations}

The inverse DFT matrix $\mathbf{D}^{-1}$ is explicitly defined using the complex exponential basis functions as follows:
\begin{equation}
\mathbf{D}^{-1} =
\begin{bmatrix}
1 & e^{i\omega t_0} & e^{i 2\omega t_0} & \dots & e^{-i\omega t_0} \\
1 & e^{i\omega t_1} & e^{i 2\omega t_1} & \dots & e^{-i\omega t_1} \\
\vdots & \vdots & \vdots & \ddots & \vdots \\
1 & e^{i\omega t_{2N_\text{H}}} & e^{i 2\omega t_{2N_\text{H}}} & \dots & e^{-i\omega t_{2N_\text{H}}}
\end{bmatrix}.
\label{eq:D_matrix}
\end{equation}

Taking the analytical derivative of the Fourier series (Eq.~\ref{eq:fourier_series}) with respect to physical time inherently introduces a multiplication by $i k \omega$ in the frequency domain. By mapping this operation back to the collocation nodes, the exact discrete temporal differentiation is formulated. Taking $\mathbf{W}^*$ as an example:

\begin{equation}
\frac{\partial \mathbf{W}^*}{\partial t} = \mathbf{D}^{-1} (i \omega \mathbf{K}) \widehat{\mathbf{W}} = i \omega \mathbf{D}^{-1} \mathbf{K} \mathbf{D}\, \mathbf{W}^* = \omega \mathbf{E}\, \mathbf{W}^*,
\label{eq:derivative_derivation}
\end{equation}
where $\mathbf{K} = \text{diag}(0, 1, \dots, N_\text{H}, -N_\text{H}, \dots, -1)$ is the diagonal wave-number matrix, and $\mathbf{E} = i \mathbf{D}^{-1} \mathbf{K} \mathbf{D}$ represents the spectral source operator that couples all temporal collocation points. Applying this exact differentiation uniformly to both the macroscopic and microscopic variables yields:
\begin{subequations}
\label{eq:spectral_derivative}
\begin{equation}
\frac{\partial \mathbf{W}^*}{\partial t} = \omega \mathbf{E}\, \mathbf{W}^*,
\label{eq:deriv_W}
\end{equation}
\begin{equation}
\frac{\partial \mathbf{h}^*}{\partial t} = \omega \mathbf{E}\, \mathbf{h}^*,
\label{eq:deriv_h}
\end{equation}
\begin{equation}
\frac{\partial \mathbf{b}^*}{\partial t} = \omega \mathbf{E}\, \mathbf{b}^*.
\label{eq:deriv_b}
\end{equation}
\end{subequations}

Remarkably, despite the presence of the imaginary unit $i$ in its derivation, the spectral differentiation matrix $\mathbf{E}$ reduces to a purely real-valued, skew-symmetric form when an odd number of collocation points $N_\text{T}$ is employed. This skew-symmetry ensures that the temporal discretization is non-dissipative. The individual elements of $\mathbf{E}$ are analytically given by:

\begin{equation}
E_{nm} =
\begin{cases}
\dfrac{1}{2}(-1)^{n-m}\csc\!\left(\dfrac{\pi(n-m)}{N_\text{T}}\right), & n\neq m\\[8pt]
0, & n=m
\end{cases}.
\label{eq:D_explicit}
\end{equation}

By applying the spectral source operator to the governing equations, the original unsteady problem is transformed into a system of coupled quasi-steady equations at each discrete temporal collocation point $n = 0, 1, \dots, 2N_\text{H}$. For the macroscopic governing equations, the resulting harmonic balance forms are given by:
\begin{equation}
\nabla_{\mathbf{x}} \cdot \mathbf{F}_n + \omega \sum_{m=0}^{N_\text{T}-1} E_{nm}\, \mathbf{W}_m = 0,
\label{eq:hb_macro}
\end{equation}
where $\mathbf{F}_n$ denotes the macroscopic spatial flux. Correspondingly, the harmonic balance equations for the reduced distribution functions are formulated as:
\begin{equation}
\mathbf{u}\cdot\nabla_{\mathbf{x}} h_n - \frac{h^{+}_n - h_n}{\tau_n} + \omega \sum_{m=0}^{N_\text{T}-1} E_{nm}\, h_m = 0,
\label{eq:hb_micro_h}
\end{equation}
\begin{equation}
\mathbf{u}\cdot\nabla_{\mathbf{x}} b_n - \frac{b^{+}_n - b_n}{\tau_n} + \omega \sum_{m=0}^{N_\text{T}-1} E_{nm}\, b_m = 0.
\label{eq:hb_micro_b}
\end{equation}

In the present time-domain collocation formulation, the nonlinear kinetic terms are evaluated directly at each temporal collocation point. Specifically, the macroscopic state $\mathbf{W}_n$ is first obtained from the moments of the distribution functions, from which $h_n^+$, $b_n^+$, and $\tau_n$ are evaluated using the  nonlinear Shakhov model. Therefore, no explicit frequency-domain convolution of the nonlinear collision term is required.
This pointwise treatment does not remove the nonlinear interaction among temporal harmonics. The nonlinear Shakhov model can generate harmonic components higher than those retained in the finite HB representation. Consequently, the temporal accuracy of the HB solution
depends on the selected harmonic truncation $N_{\mathrm H}$, whose adequacy should be examined numerically through harmonic-resolution studies. The physical-time derivative, on the other hand, is represented by the linear spectral operator
$\omega\mathbf E$, which couples the solution values at all temporal collocation points.

To solve the coupled harmonic-balance equations, an auxiliary pseudo-time $\Gamma$ is introduced. The target periodic solution is obtained as the fixed point of an auxiliary pseudo-time system constructed from the complete HB residual:
\begin{subequations}
\label{eq:pseudotime_hb}

\begin{equation}
\frac{\partial \mathbf{W}_n}{\partial \Gamma}
+\nabla_{\mathbf{x}}\cdot\mathbf{F}_n
+\omega\sum_{m=0}^{N_{\mathrm T}-1}
E_{nm}\mathbf{W}_m
=0,
\end{equation}

\begin{equation}
\frac{\partial h_n}{\partial \Gamma}
+\mathbf{u}\cdot\nabla_{\mathbf{x}}h_n
-\frac{h_n^+-h_n}{\tau_n}
+\omega\sum_{m=0}^{N_{\mathrm T}-1}
E_{nm}h_m
=0,
\end{equation}

\begin{equation}
\frac{\partial b_n}{\partial \Gamma}
+\mathbf{u}\cdot\nabla_{\mathbf{x}}b_n
-\frac{b_n^+-b_n}{\tau_n}
+\omega\sum_{m=0}^{N_{\mathrm T}-1}
E_{nm}b_m
=0.
\end{equation}

\end{subequations}

For convenience, the HB spectral source associated with a generic variable $q_n$ is denoted by
\begin{equation}
\mathcal{S}^{\mathrm{HB}}_{q,n}
=
\omega\sum_{m=0}^{N_{\mathrm T}-1}E_{nm}q_m.
\label{eq:hb_source}
\end{equation}

Formally, assuming that $\tau_n$ is locally frozen over the interface evolution interval, if the HB spectral source were incorporated directly into the characteristic evolution, the solution of the complete pseudo-time kinetic equation would contain an additional
source contribution:
\begin{equation}
\begin{aligned}
h_n(\mathbf{x}_f,\Gamma,\mathbf{u})
={}&
e^{-\Gamma/\tau_n}
h_{n,0}(\mathbf{x}_f-\mathbf{u}\Gamma,\mathbf{u})
\\
&+
\frac{1}{\tau_n}
\int_0^\Gamma
h_n^+(\mathbf{x}',\Gamma',\mathbf{u})
e^{-(\Gamma-\Gamma')/\tau_n}
\,d\Gamma'
\\
&-
\int_0^\Gamma
\mathcal{S}^{\mathrm{HB}}_{h,n}
(\mathbf{x}',\Gamma',\mathbf{u})
e^{-(\Gamma-\Gamma')/\tau_n}
\,d\Gamma',
\end{aligned}
\label{eq:full_characteristic}
\end{equation}
where
\[
\mathbf{x}'
=
\mathbf{x}_f-\mathbf{u}(\Gamma-\Gamma').
\]

In the present implementation, however, the HB spectral source is not embedded into the local characteristic evolution used for UGKS flux construction. Instead, the pseudo-time residual is decomposed into a local transport--collision contribution and a global HB spectral source. The former is used to construct the multiscale interface flux, whereas the latter is retained explicitly at the cell level in the pseudo-time update.
This source-separated treatment is adopted because the HB source $\mathcal{S}^{\mathrm{HB}}_{h,n}$ couples all temporal collocation states. Incorporating it directly into the characteristic solution would make the interface-flux evaluation at a given temporal collocation point depend on the microscopic states at all other collocation points. Such a construction would introduce global temporal coupling directly into the interface-flux evaluation and substantially increase the associated algebraic complexity.

It should be emphasized that this decomposition is introduced only for the numerical construction of the pseudo-time iteration. The HB spectral source is not discarded from the governing residual. Since $\Gamma$ is an auxiliary iteration variable rather than the physical time, the objective is the converged fixed point of the complete HB residual rather than an accurate transient evolution in $\Gamma$. Accordingly, the local auxiliary evolution used exclusively for interface-flux construction at the temporal collocation point $t_n$ is defined as
\begin{equation}
\frac{\partial h_n^{\mathrm{loc}}}{\partial\Gamma}
+
\mathbf{u}\cdot\nabla_{\mathbf{x}}h_n^{\mathrm{loc}}
=
\frac{
h_n^{+,\mathrm{loc}}-h_n^{\mathrm{loc}}
}{
\tau_n
}.
\label{eq:local_ugks_evolution}
\end{equation}

The corresponding local integral solution at the interface is
\begin{equation}
\begin{aligned}
h_n^{\mathrm{loc}}(\mathbf{x}_f,\Gamma,\mathbf{u})
={}&
\frac{1}{\tau_n}
\int_0^\Gamma
h_n^{+,\mathrm{loc}}
(\mathbf{x}',\Gamma',\mathbf{u})
e^{-(\Gamma-\Gamma')/\tau_n}
\,d\Gamma'
\\
&+
e^{-\Gamma/\tau_n}
h_{n,0}(\mathbf{x}_f-\mathbf{u}\Gamma,\mathbf{u}).
\end{aligned}
\label{eq:integral_solution}
\end{equation}

Equation~(\ref{eq:integral_solution}) is therefore the integral solution of the local transport--collision operator used for UGKS flux construction, rather than the characteristic solution of the complete HB pseudo-time equation~(\ref{eq:pseudotime_hb}). The HB
spectral source omitted from this local interface evolution is subsequently restored explicitly in the cell-centered residual. For notational simplicity, the superscript ``loc'' is omitted hereafter in the interface-flux expressions.

To achieve second-order spatial accuracy, the equilibrium state is expanded locally around the cell interface as $h_n^+ = h_{n,0}^+[1 + a_n x_n + a_t x_t + A \Gamma]$, where the coefficients $a_n, a_t$ and $A$ are derived from the Taylor expansion of the Maxwellian distribution. Meanwhile, the initial non-equilibrium distribution $h_{n,0}$ is reconstructed using a monotonicity-preserving limiter. It is worth noting that the macroscopic variables at the interface, which determine the initial equilibrium state $h_{n,0}^+$, are obtained through kinetic averaging of the reconstructed left and right states. By performing a time-averaging integration of
Eq.~(\ref{eq:integral_solution}) over the face-level pseudo-time
interval $[0,\Delta\Gamma_f]$, the time-averaged interface distribution
is obtained analytically. The resulting coefficients are defined as
\begin{equation}
\begin{aligned}
C_1 &= \Delta\Gamma_f
-\tau_n\left(1-e^{-\Delta\Gamma_f/\tau_n}\right), \\
C_2 &= 2\tau_n^2\left(1-e^{-\Delta\Gamma_f/\tau_n}\right)
-\tau_n\Delta\Gamma_f
-\tau_n\Delta\Gamma_f e^{-\Delta\Gamma_f/\tau_n}, \\
C_3 &= \frac{\Delta\Gamma_f^2}{2}
-\tau_n\Delta\Gamma_f
+\tau_n^2\left(1-e^{-\Delta\Gamma_f/\tau_n}\right), \\
C_4 &= \tau_n\left(1-e^{-\Delta\Gamma_f/\tau_n}\right), \\
C_5 &= \tau_n\Delta\Gamma_f e^{-\Delta\Gamma_f/\tau_n}
-\tau_n^2\left(1-e^{-\Delta\Gamma_f/\tau_n}\right).
\end{aligned}
\label{eq:C_coeffs}
\end{equation}

The pseudo-time-averaged microscopic flux of the reduced distribution
$h$ at face $f$, temporal collocation point $t_n$, and discrete
velocity point $\alpha$ is written as
\begin{equation}
\begin{aligned}
\mathcal{H}_{f,n,\alpha}
=
\frac{u_{\alpha,\mathrm n}}{\Delta\Gamma_f}
\Big[
& C_1 h^{+}_{f,n,\alpha}
+ C_2
\left(
u_{\alpha,\mathrm n} a_{\mathrm n}
+
u_{\alpha,\mathrm t} a_{\mathrm t}
\right)
h^{+}_{f,n,0,\alpha}
\\
&+ C_3 A h^{+}_{f,n,0,\alpha}
+ C_4 h_{f,n,0,\alpha}
\\
&+ C_5
\left(
u_{\alpha,\mathrm n}\sigma_{\mathrm n}^{h}
+
u_{\alpha,\mathrm t}\sigma_{\mathrm t}^{h}
\right)
\Big].
\end{aligned}
\label{eq:micro_flux}
\end{equation}

Here, $u_{\alpha,\mathrm n}$ and $u_{\alpha,\mathrm t}$ denote the
normal and tangential components of the discrete molecular velocity
$\mathbf{u}_\alpha$, respectively.

The same local integral construction is applied to the reduced
distribution $b$, yielding the corresponding microscopic interface
flux $\mathcal{B}_{f,n,\alpha}$. The macroscopic conservative flux is
then obtained by taking velocity moments of the two reduced
microscopic fluxes:
\begin{equation}
\mathbf{F}_{f,n}
=
\sum_{\alpha} w_\alpha
\begin{bmatrix}
\mathcal{H}_{f,n,\alpha}
\\[3pt]
\mathbf{u}_\alpha\,\mathcal{H}_{f,n,\alpha}
\\[3pt]
\dfrac{1}{2}
\left(
|\mathbf{u}_\alpha|^2
\mathcal{H}_{f,n,\alpha}
+
\mathcal{B}_{f,n,\alpha}
\right)
\end{bmatrix}.
\label{eq:macro_flux}
\end{equation}

\subsection{Boundary conditions}

For solid walls, the isothermal fully diffuse reflection boundary condition is employed. In the context of the present time-domain harmonic balance framework, the solid boundaries may be subjected to periodic harmonic excitations. Accordingly, the prescribed wall velocity $\mathbf{U}_w$ and temperature $T_w$ are explicitly evaluated at each discrete temporal collocation point $t_n$:
\begin{equation}
\mathbf{U}_w^n = \overline{\mathbf{U}}_w + \Re\left( \widehat{\mathbf{U}}_w e^{i \omega t_n} \right),
\label{eq:bc_vel}
\end{equation}
\begin{equation}
T_w^n = \overline{T}_w + \Re\left( \widehat{T}_w e^{i \omega t_n} \right),
\label{eq:bc_temp}
\end{equation}
where $\overline{\mathbf{U}}_w$ and $\overline{T}_w$ denote the time-averaged wall velocity and temperature, respectively. The terms $\widehat{\mathbf{U}}_w$ and $\widehat{T}_w$ represent the amplitudes of the harmonic excitation. Here, $\Re(\cdot)$ extracts the real physical value to explicitly determine the instantaneous boundary conditions at each specific time level $t_n$.

To preserve uniform second-order spatial accuracy and the multiscale characteristics up to the solid boundary $\mathbf{x}_w$, the multiscale boundary condition proposed by Chen et al. \cite{chen2013comparative} is adopted. Taking the reduced distribution $h_n$ as an example, the distribution function at the wall interface is constructed as:
\begin{equation}
h_n(\mathbf{x}_w, \mathbf{u}) =
\begin{cases}
\rho_w^n\, H_w(\mathbf{u}; \mathbf{U}_w^n, T_w^n), & \mathbf{u}\cdot\mathbf{n}_{\text{in}} > 0, \\[4pt]
h_n^{\text{interior}}(\mathbf{x}_w, \mathbf{u}), & \mathbf{u}\cdot\mathbf{n}_{\text{in}} \leq 0
\end{cases},
\label{eq:bc_diffuse}
\end{equation}
where $\mathbf{n}_{\text{in}}$ is the normal vector pointing from the wall into the flow field. Here, $H_w$ is the normalized baseline Maxwellian completely determined by the instantaneous wall state $(\mathbf{U}_w^n, T_w^n)$.

Crucially, for the incoming particles ($\mathbf{u}\cdot\mathbf{n}_{\text{in}} \leq 0$), the incident distribution $h_n^{\text{interior}}$ is not a simple zero-order extrapolation from the adjacent cell center. Instead, it is rigorously evaluated using the same characteristics-based integral solution (Eq.~\ref{eq:integral_solution}) from the interior domain to the wall. With the incident distribution obtained, the instantaneous wall density $\rho_w^n$ for the reflected particles ($\mathbf{u}\cdot\mathbf{n}_{\text{in}} > 0$) is evaluated by enforcing the strict impermeability condition. This condition dictates a zero net mass flux across the solid boundary at any temporal collocation point $n$:
\begin{equation}
\rho_w^n = -\frac{
\displaystyle\sum_{\mathbf{u}\cdot\mathbf{n}_{\text{in}} < 0} w_{\alpha}\, (\mathbf{u}\cdot\mathbf{n}_{\text{in}})\, h_n^{\text{interior}}
}{
\displaystyle\sum_{\mathbf{u}\cdot\mathbf{n}_{\text{in}} > 0} w_{\alpha}\, (\mathbf{u}\cdot\mathbf{n}_{\text{in}})\, H_w
}.
\label{eq:rho_wall}
\end{equation}
Once $\rho_w^n$ is obtained, the complete microscopic distribution functions $(h_n, b_n)$ at the boundary are fully determined. Subsequently, the macroscopic boundary fluxes $\mathbf{F}_n$ are computed by taking the conservative moments of Eq.~(\ref{eq:bc_diffuse}) over the discrete velocity space, thereby providing the exact boundary flux closure for the macroscopic harmonic balance equations.

\subsection{Finite-volume discretization and pseudo-time marching}
\label{sec:pseudotime_update}

Based on the source-separated construction described above, the UGKS interface flux and the HB spectral source enter the finite-volume residual through two distinct contributions. The interface flux is evaluated locally at each temporal collocation point using Eq.~(\ref{eq:integral_solution}), while the global temporal coupling is introduced through the cell-centered HB source $\mathcal{S}^{\mathrm{HB}}_{q,n}$. Applying a cell-centered finite-volume discretization over the control volume $\Omega_i$, the macroscopic update at the $n$-th temporal collocation point is written as
\begin{equation}
\frac{
\mathbf{W}_{i,n}^{k+1}-\mathbf{W}_{i,n}^{k}
}{
\Delta\Gamma_i
}
+
\frac{1}{|\Omega_i|}
\sum_{f\in\partial\Omega_i}
|L_f|\,\mathbf{F}_{f,n}^{k}
+
\omega
\sum_{m=0}^{N_{\mathrm T}-1}
E_{nm}\mathbf{W}_{i,m}^{k}
=0,
\label{eq:fv_W}
\end{equation}
where $\Delta\Gamma_i$ denotes the local pseudo-time step, $|L_f|$ is the face length, and $\mathbf{F}_{f,n}^{k}$ is the macroscopic UGKS interface flux obtained from the local
transport--collision evolution described above.

For the evolution of the microscopic reduced distribution functions $(h,b)$, the convective flux and the HB spectral source are treated explicitly in pseudo-time, while the stiff relaxation term is treated semi-implicitly. Taking $h$ as an example, the update equation is written as
\begin{equation}
h_{i,n}^{k+1}
=
\frac{
h_{i,n}^{k}
-
\frac{\Delta\Gamma_i}{|\Omega_i|}
\sum_f |L_f|\,\mathcal{H}_{f,n}^{k}
-
\Delta\Gamma_i\,
\omega\sum_mE_{nm}h_{i,m}^{k}
+
\frac{\Delta\Gamma_i}{2}
\left(
\frac{h_{i,n}^{+,k+1,*}}{\tau_{i,n}^{k+1}}
+
\frac{h_{i,n}^{+,k}-h_{i,n}^{k}}{\tau_{i,n}^{k}}
\right)
}{
1+\frac{\Delta\Gamma_i}{2\tau_{i,n}^{k+1}}
},
\label{eq:fv_h}
\end{equation}
where $\mathcal{H}_{f,n}^{k}$ denotes the microscopic UGKS interface flux evaluated from Eq.~(\ref{eq:micro_flux}). The HB spectral source is treated explicitly through $\omega\sum_mE_{nm}h_{i,m}^{k}$, whereas the stiff relaxation term is treated semi-implicitly.  The superscript $*$ indicates that the Shakhov target state is
evaluated using the updated conservative variables and an explicitly
estimated heat flux, rather than the fully updated distribution
functions:
\begin{equation}
h_{i,n}^{+,k+1,*}
=
\mathcal{S}_h
\left(
\mathbf{W}_{i,n}^{k+1},
\widetilde{\mathbf{q}}_{i,n}^{\,k+1}
\right),
\label{eq:shakhov_target_predictor}
\end{equation}
where $\mathcal{S}_h$ denotes the reduced Shakhov target distribution.
The corresponding target for $b$ is constructed in the same manner.

The consistency of this source-separated pseudo-time iteration with the target HB system can be seen directly from its converged fixed point. When pseudo-time convergence is reached,
\[
h_{i,n}^{k+1}=h_{i,n}^{k}=h_{i,n},
\qquad
h_{i,n}^{+,k+1,*}=h_{i,n}^{+,k}=h_{i,n}^{+},
\qquad
\tau_{i,n}^{k+1}=\tau_{i,n}^{k}=\tau_{i,n},
\]
and Eq.~(\ref{eq:fv_h}) reduces to
\begin{equation}
\frac{1}{|\Omega_i|}
\sum_f |L_f|\,\mathcal{H}_{f,n}
-
\frac{h_{i,n}^{+}-h_{i,n}}{\tau_{i,n}}
+
\omega\sum_mE_{nm}h_{i,m}
=0,
\label{eq:hb_fixed_point}
\end{equation}
which is precisely the finite-volume residual form of the microscopic HB equation~(\ref{eq:hb_micro_h}). Therefore, although the HB spectral source is not embedded in the local UGKS characteristic solution used for interface-flux construction, it remains an integral part of the converged discrete HB system.

The allowable local pseudo-time step $\Delta\Gamma_i$ is estimated using a generalized Courant--Friedrichs--Lewy (CFL) condition that
accounts for both convective transport and the stiffness introduced by the HB temporal operator:
\begin{equation}
\Delta\Gamma_i
=
\frac{
\mathrm{CFL}
}{
\nu_{\mathrm{conv},i}
+
\omega N_{\mathrm H}
}.
\label{eq:dt_hb}
\end{equation}

Here, $\omega N_{\mathrm H}$ is the spectral radius of the scaled temporal differentiation operator $\omega\mathbf E$, while $\nu_{\mathrm{conv},i}$ denotes the local estimate of the convective spectral radius,
\begin{equation}
\nu_{\mathrm{conv},i}
=
\frac{
\Delta y_i\,\widetilde{u}_{\max,i}
+
\Delta x_i\,\widetilde{v}_{\max,i}
}{
|\Omega_i|
},
\label{eq:nu_conv}
\end{equation}
with
\[
\widetilde{u}_{\max,i}
=
\max(|u_{\max}|,|U_i|)+c_i,
\qquad
\widetilde{v}_{\max,i}
=
\max(|v_{\max}|,|V_i|)+c_i,
\]
where
$c_i=\sqrt{\gamma RT_i}$ is the local sound speed.

Since $\Gamma$ is an auxiliary pseudo-time variable introduced solely for iterative convergence, no physical temporal-accuracy requirement is imposed on its advancement. A local time-stepping strategy is
therefore adopted, allowing each control volume to advance with its own pseudo-time step. For a face $f$ shared by left and right cells, however, the kinetic integral solution is evolved with a face-level step $\Delta\Gamma_f$ assembled from the two neighboring cell values:
\begin{equation}
\Delta\Gamma_f =
\begin{cases}
\min(\Delta\Gamma_L,\Delta\Gamma_R),\\[6pt]
\dfrac{2\Delta\Gamma_L\Delta\Gamma_R}{\Delta\Gamma_L+\Delta\Gamma_R}
\end{cases},
\label{eq:dt_face}
\end{equation}
where the first and second expressions correspond to the minimum and harmonic-average strategies, respectively. Unless otherwise specified, the minimum strategy is employed in the present work. The UGKS integral solution and the time-integrated flux at the cell interface are first evaluated over the interval $[0,\Delta\Gamma_f]$. The resulting face contribution is then divided by $\Delta\Gamma_f$ to yield a time-averaged conservative flux rate, which is subsequently multiplied by the local pseudo-time step $\Delta\Gamma_i$ during the updates in Eqs.~(\ref{eq:fv_W}) and (\ref{eq:fv_h}). In this manner, the interface evolution remains physically consistent with the local kinetic scales of the adjacent cells, while each control volume retains the efficiency of LTS. Specifically, under low-frequency excitations, $\Delta\Gamma_i$ is governed mainly by the standard convective CFL condition, whereas under high-frequency conditions, it is increasingly restricted by the stiffness of the time spectral source term. Ultimately, LTS improves the pseudo-time convergence because slowly varying regions are no longer forced to march with the smallest step dictated by the stiffest cell, while the face-level selection in Eq.~(\ref{eq:dt_face}) preserves a consistent evolution interval in the UGKS flux construction.

\begin{figure}[htbp]
\centering
\includegraphics[width=1\textwidth]{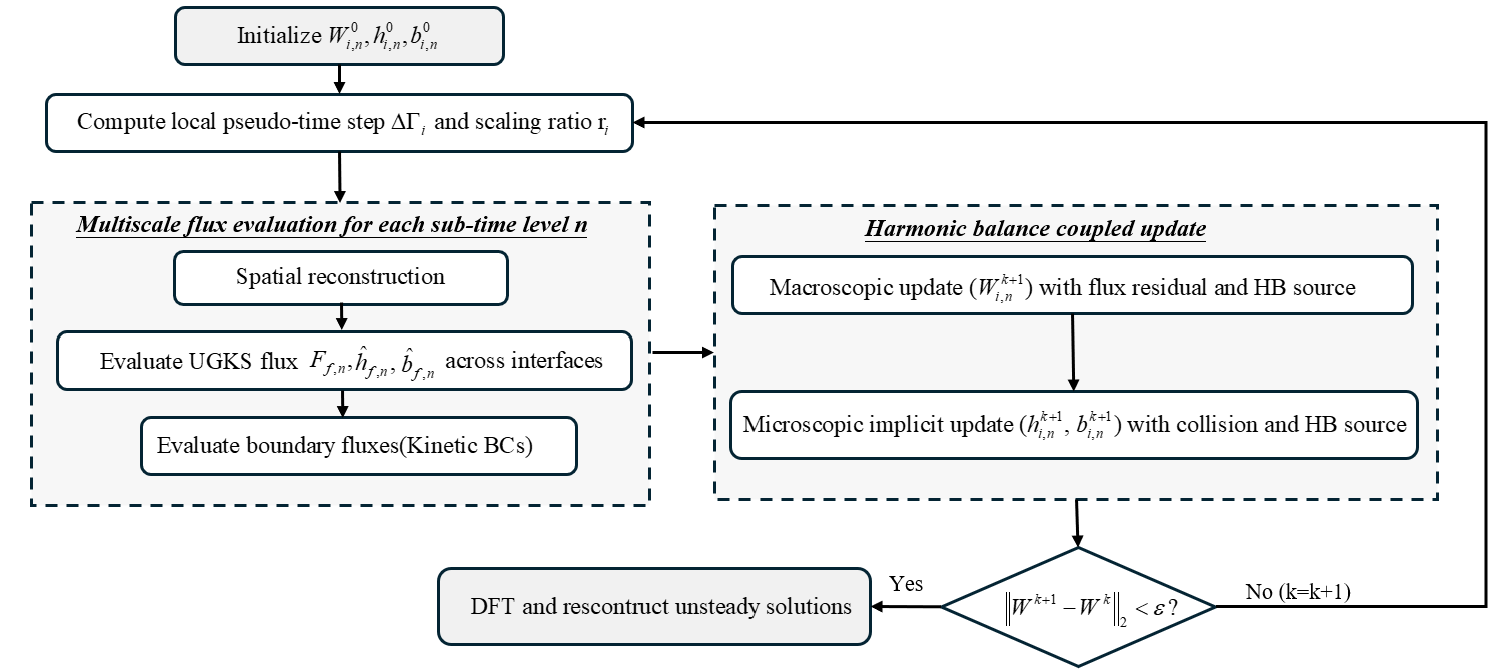}
\caption{Flowchart of the pseudo-time marching procedure for the HB-UGKS solver.}
\label{fig:algorithm_flow}
\end{figure}

Overall, the proposed HB-UGKS framework inherently incorporates three distinct time scales. The first is the macroscopic physical period $T$, represented by $N_\text{T}$ collocation points over one period. The second is the local kinetic evolution time $[0,\Delta\Gamma_f]$, explicitly utilized during the interface flux evaluation to capture the multiscale characteristics of the flow. The third is the pseudo-time $\Gamma$ for iteration, which relaxes the coupled harmonic balance system toward convergence. While the interface flux evaluation remains strictly localized at each collocation point, the time spectral operator introduces a global temporal coupling across the entire period. For clarity, the overall pseudo-time marching procedure is illustrated in Fig.~\ref{fig:algorithm_flow}.

\section{Results and discussion}

The three benchmarks follow a progression from two linear shear-driven problems to finite-amplitude nonlinear thermal forcing. The first two linear cases—the planar cavity and the eccentric annulus—establish the baseline response and dynamic wall damping, where the flow remains dominated by the fundamental excitation frequency, while the thermally driven cavity assesses the required harmonic component beyond the linear regime. Each subsection closes with an HB--TD efficiency assessment, so that the computational speedup can be evaluated alongside the accuracy for the same case.

\subsection{Oscillatory shear-driven cavity flow}
\label{sec:shear_cavity}

\subsubsection{Numerical configuration and harmonic verification}

The two-dimensional shear-driven oscillatory cavity flow serves as a standard benchmark for verifying unsteady kinetic solvers. The computational domain is a rectangular cavity defined by $(x, y) \in [0, L] \times [0, H]$, with an aspect ratio of $A = L/H$. The bottom, left, and right walls are stationary and maintained at a constant reference temperature $T_0$. The flow is driven by the top lid ($y=H$), which oscillates harmonically in its own plane with a prescribed velocity:

\begin{equation}
U_w(t) = \overline{U}_w + U_a \cos(\omega t),
\label{eq:lid_velocity}
\end{equation}

\begin{equation}
V_w(t) = 0,
\label{eq:lid_velocityV}
\end{equation}
where $\overline{U}_w$ denotes the time-averaged horizontal velocity of the lid, and $U_a$ is the oscillation amplitude. In the present simulations, the time-averaged  horizontal  velocity of lid is set to zero ($\overline{U}_w = 0$), and a sufficiently small oscillation amplitude of $U_a = 0.0015\sqrt{2RT_0}$ is adopted to ensure the flow closely approximates the linear regime. This configuration is consistent with the linearized Boltzmann equation framework proposed by Wu et al.~\cite{wu2014oscillatory}.

The flow regime is governed by two fundamental dimensionless parameters. The Knudsen number ($\text{Kn}$), defined based on the cavity height $H$, is expressed as:
\begin{equation}
Kn = \frac{\lambda_{mfp}}{H} = \frac{4\alpha(5-2\omega_{VSS})(7-2\omega_{VSS})}{5(\alpha+1)(\alpha+2)} \frac{\mu(T_0)}{\rho_0 H \sqrt{2\pi R T_0}},
\label{eq:def_Kn}
\end{equation}
where $\lambda_{mfp}$ is the molecular mean free path, $\rho_0$ is the initial uniform density, and $\mu(T_0)$ is the reference dynamic viscosity. To accurately model Argon gas, the Variable Soft Sphere (VSS) collision model is employed. The dynamic viscosity follows a standard power-law temperature dependence with an index of $\omega_{VSS} = 0.81$ and an angular scattering parameter of $\alpha = 1.0$. The Strouhal number ($\text{St}$), which characterizes the ratio of the molecular collision time \(\tau_0\) to the oscillation time scale \(1/\omega\)., is defined as:
\begin{equation}
St = \frac{\omega H}{\sqrt{2RT_0}}.
\label{eq:def_St}
\end{equation}

For the standard square cavity ($A = 1$), the physical space is discretized using a $90 \times 90$ non-uniform structured mesh. To accurately resolve the near-wall multiscale gradients and Knudsen layer effects, the mesh is progressively refined towards all physical boundaries with a minimum grid spacing of $\Delta s_{\min} = 0.001H$ adjacent to the walls. For cavities with different aspect ratios, the number of cells in the horizontal direction is scaled proportionally while maintaining the identical vertical resolution: $N_x = 180$ for $A = 2$, and $N_x = 45$ for $A = 0.5$.

In the velocity space, the continuous microscopic velocity is truncated at $[-4, 4]^2$ (normalized by $\sqrt{2RT_0}$). A regime-dependent discretization strategy is adopted to balance accuracy and computational efficiency. For the near-continuum and slip regimes ($Kn \leq 0.1$), a $28 \times 28$ Gauss-Hermite quadrature is employed, whereas a $64 \times 64$ Newton-Cotes rule is utilized for the highly rarefied regimes ($Kn \geq 1.0$). In these standard cases, the reference viscosity $\mu(T_0)$ is explicitly dictated by the prescribed Knudsen number via Eq.~(\ref{eq:def_Kn}). Conversely, for the specific continuum limit cases, a $16 \times 16$ uniform discrete velocity grid is sufficient, and the reference viscosity is instead determined by the macroscopic Reynolds number as:
\begin{equation}
\mu(T_0) =\frac{ \rho_0 \overline{U}_w L}{Re}.
\label{eq:def_Re_mu0}
\end{equation}

As described previously, the conservative macroscopic variables at each temporal collocation point $n$ are directly obtained from the velocity-space moments of the distribution function $h_n$. Furthermore, the local shear stress, which is essential for characterizing the aerodynamic viscous response, can be evaluated as:
\begin{equation}
\tau_{xy,n} = \sum_{\alpha} w_\alpha (u_\alpha - U_n)(v_\alpha - V_n) h_n \label{eq:shear_stress_eval}.
\end{equation}

Once the pseudo-time marching of the HB-UGKS converges, the periodic steady-state solutions (including both macroscopic variables and microscopic distribution functions) at the $N_\text{T}$ temporal collocation points are obtained. Subsequently, a discrete Fourier transform is applied as a post-processing step to extract the Fourier coefficients for each variable. The $k$-th order complex Fourier coefficient $\widehat{q}_k(\mathbf{x})$ is defined as:
\begin{equation}
\widehat{q}_k(\mathbf{x}) = \frac{1}{N_\text{T}} \sum_{n=0}^{N_\text{T}-1} q_n(\mathbf{x}) e^{-i k \omega t_n},
\label{eq:dft_general}
\end{equation}
where $q_n(\mathbf{x})$ represents any variable of interest at the $n$-th temporal collocation point. The coefficient $\widehat{q}_0$ gives the time-averaged component. For $k\geq1$, $\widehat{q}_k$ denotes the positive-frequency coefficient in the
two-sided Fourier representation. For comparison with conventional harmonic amplitudes, we define the corresponding one-sided complex amplitude as
\begin{equation}
q_k^{c}(\mathbf{x}) = 2\widehat{q}_k(\mathbf{x}),
\qquad k\geq1.
\label{eq:complex_amplitude}
\end{equation}
Accordingly, the real physical signal may be written as
\begin{equation}
q(\mathbf{x},t)
=
\widehat{q}_0(\mathbf{x})
+
\Re\left[
\sum_{k=1}^{N_\text{H}}
q_k^{c}(\mathbf{x})e^{ik\omega t}
\right].
\label{eq:physical_reconstruction}
\end{equation}
Hereafter, the reported harmonic amplitudes are
based on
$q_k^{c}$.

\begin{figure}[htbp]
    \centering
    \includegraphics[width=0.52\linewidth]{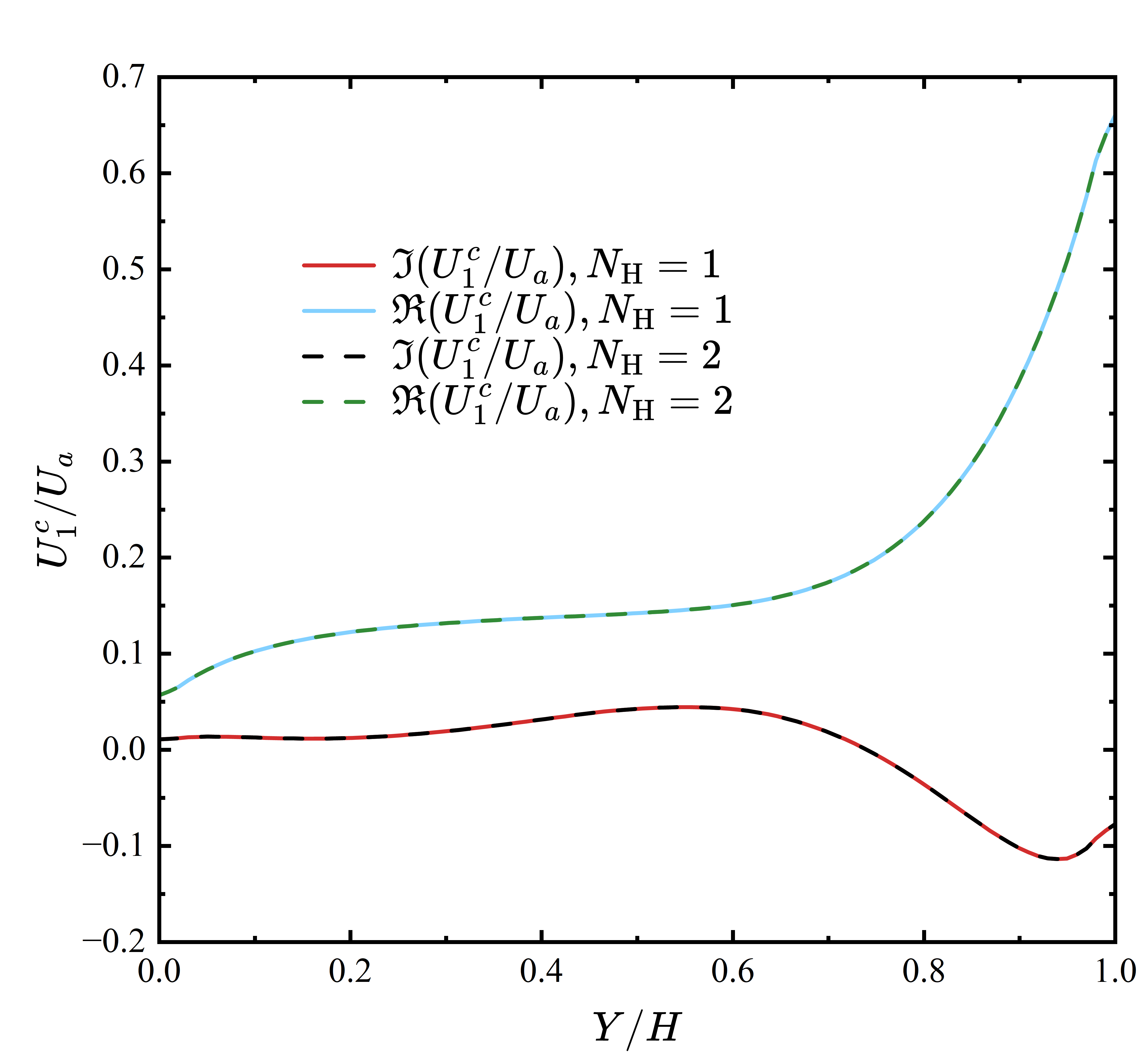}
    \caption{Harmonic independence verification: profiles of
    $\Re(U_1^{c}/U_a)$ and $\Im(U_1^{c}/U_a)$ along the vertical centerline $x=0.5L$ for $N_\text{H}=1$ and $N_\text{H}=2$
     ($Kn=0.1$, $St=2$).}
    \label{fig:HB_verification_Wu}
\end{figure}

Before proceeding to comprehensive flow investigations, the required harmonic truncation order $N_\text{H}$ is verified. Given the small excitation amplitude, the flow response is expected to be dominated by the fundamental frequency. To substantiate that $N_\text{H} = 1$ provides sufficient fidelity, simulations with $N_\text{H} = 1$ and $N_\text{H} = 2$ are compared at $Kn = 0.1$ and $St = 2$. As depicted in Fig.~\ref{fig:HB_verification_Wu}, the resulting vertical centerline velocity profiles  are virtually identical, demonstrating that higher-harmonic contributions are negligible. Accordingly, $N_\text{H} = 1$ is retained for all weakly perturbed shear-driven simulations.

\subsubsection{Gas anti-resonance and high-frequency limits}
To quantitatively evaluate the overall aerodynamic response on the
oscillating lid, the reference shear-stress scale is defined as
\begin{equation}
\tau_{\mathrm{ref}}
=
\frac{p_0 U_a}{\sqrt{2RT_0}}.
\label{eq:tau_ref}
\end{equation}

The spatially averaged fundamental amplitudes of the horizontal
velocity and shear stress are then normalized as
\begin{equation}
\widetilde{U}
=
\frac{\big|\langle U_1^{c}\rangle_{\mathrm{wall}}\big|}{U_a},
\qquad
\widetilde{\tau}_{xy}
=
\frac{\big|\langle \tau_{xy,1}^{c}\rangle_{\mathrm{wall}}\big|}
{\tau_{\mathrm{ref}}}.
\label{eq:nondim_global}
\end{equation}

where $\langle \cdot \rangle_{\text{wall}}$ denotes the spatial average along the top boundary of length $L$, evaluated as a discrete integral weighted by the local cell width: $\langle q \rangle_{\text{wall}} = \frac{1}{L} \sum_{i=1}^{N_x} q_i \Delta x_i$.

\begin{figure}[htbp]
    \centering
    \begin{subfigure}{0.48\textwidth}
        \includegraphics[width=\linewidth]{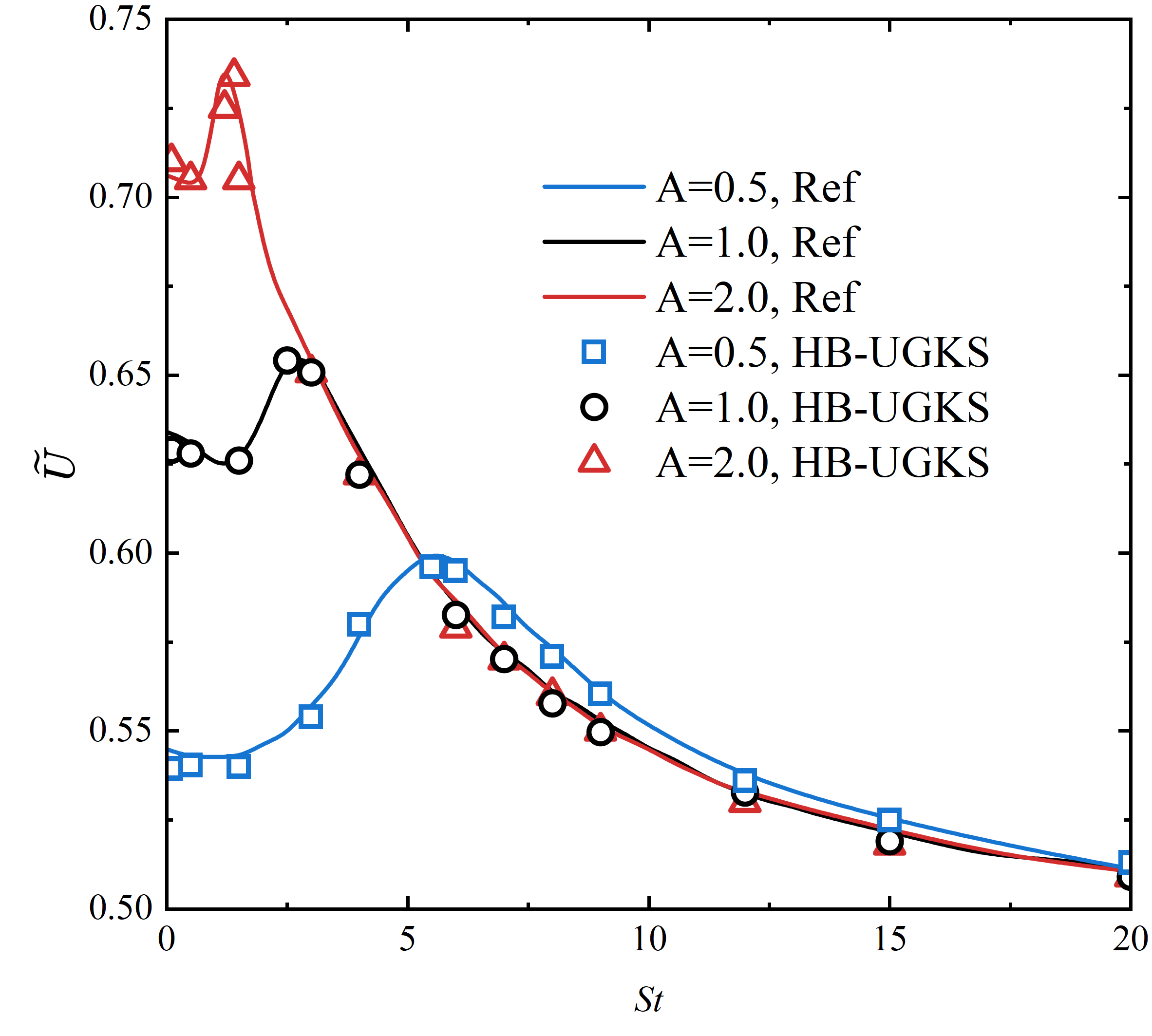}
        \caption{$\widetilde{U}$ versus $St$}
    \end{subfigure}
    \hfill
    \begin{subfigure}{0.48\textwidth}
        \includegraphics[width=\linewidth]{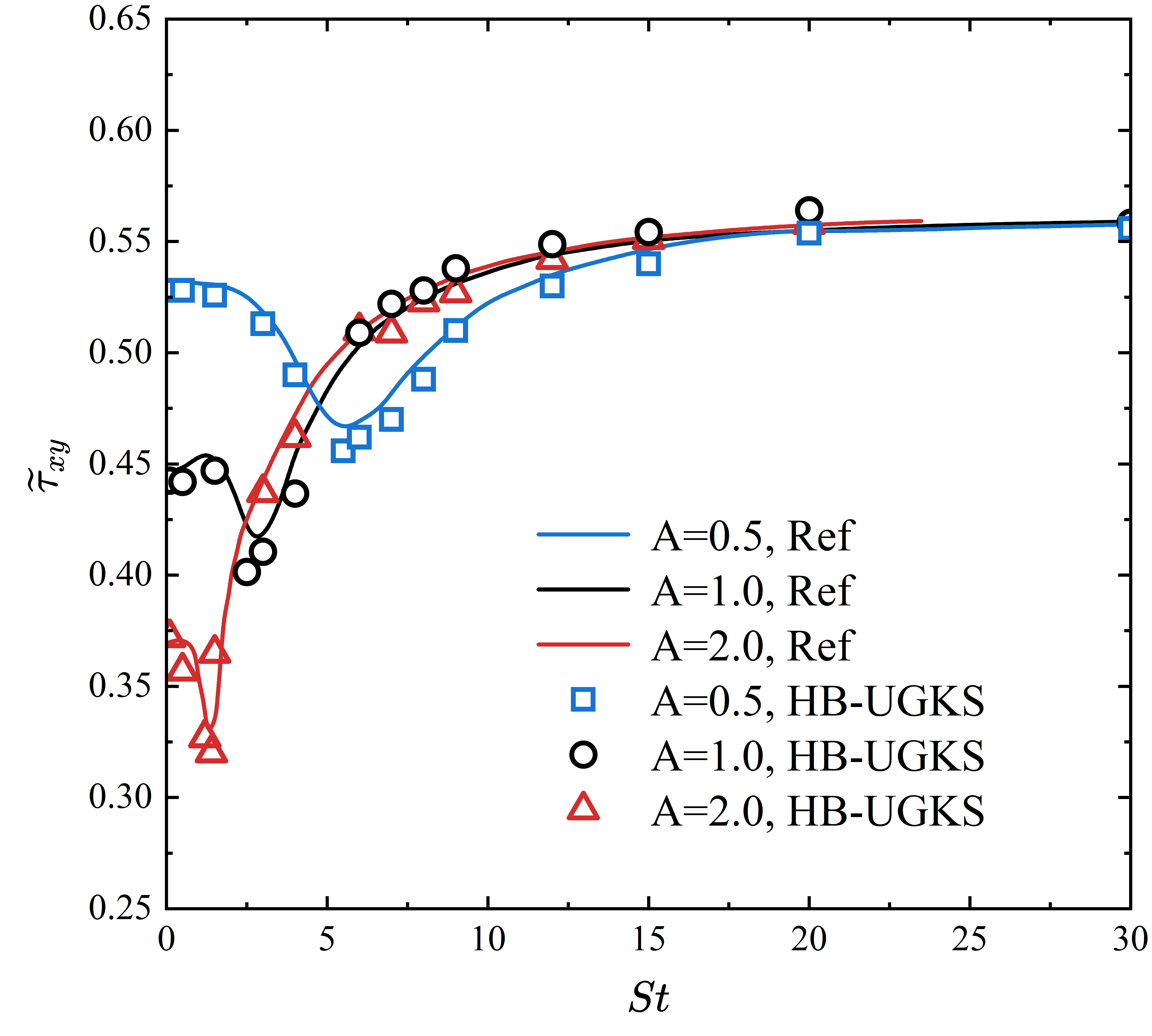}
        \caption{$\widetilde{\tau}_{xy}$ versus $St$}
    \end{subfigure}
    \caption{Spatially averaged fundamental amplitudes versus Strouhal number $St$ at $Kn = 0.1$ for various cavity aspect ratios. Symbols: HB-UGKS; solid lines: reference solutions from Wu et al.~\cite{wu2014oscillatory}.}
    \label{fig:Wu_Global_Amplitudes}
\end{figure}

Figure~\ref{fig:Wu_Global_Amplitudes} shows the evolution of $\widetilde{U}$ and $\widetilde{\tau}_{xy}$ across a broad range of Strouhal numbers for three cavity aspect ratios at $Kn = 0.1$. The HB-UGKS solutions demonstrate excellent agreement with the linearized Boltzmann reference data of Wu et al.~\cite{wu2014oscillatory} across the entire frequency spectrum. Notably, the wall shear stress exhibits a non-monotonic variation, developing a distinct local minimum that characterizes the gas anti-resonance phenomenon. Physically, this minimum originates from the destructive interference between the primary viscous wave driven by the oscillating lid and the secondary acoustic waves reflected from the stationary side walls. Because the lateral travel distance governs the phase condition for this interference, the anti-resonant minimum systematically shifts toward higher Strouhal numbers as the cavity aspect ratio $A$ decreases.

In the high-frequency limit ($St = 30$), the oscillation time scale $1/\omega$ becomes shorter than the local mean collision time in the
near-wall region. As a result, intermolecular collisions are no longer able to fully redistribute the momentum imposed by the rapidly
oscillating boundary within one local oscillation time scale. The gas response near the lid therefore approaches the collisionless half-space limit, even though the global Knudsen number remains $Kn=0.1$. Under this condition, the local distribution function
adjacent to the wall is well approximated by a superposition of an unperturbed incident half-Maxwellian and a fully accommodated reflected half-Maxwellian. Integrating this half-space distribution theoretically demonstrates that $\widetilde{U}$ and $\widetilde{\tau}_{xy}$ asymptotically approach $0.5$ and $1/\sqrt{\pi} \approx 0.5642$, respectively. The HB-UGKS precisely recovers these collisionless asymptotes across all aspect ratios, with mean relative deviations bounded well within $1.5\%$.

\begin{figure}[htbp]
    \centering
    \begin{subfigure}{0.48\textwidth}
        \includegraphics[width=\linewidth]{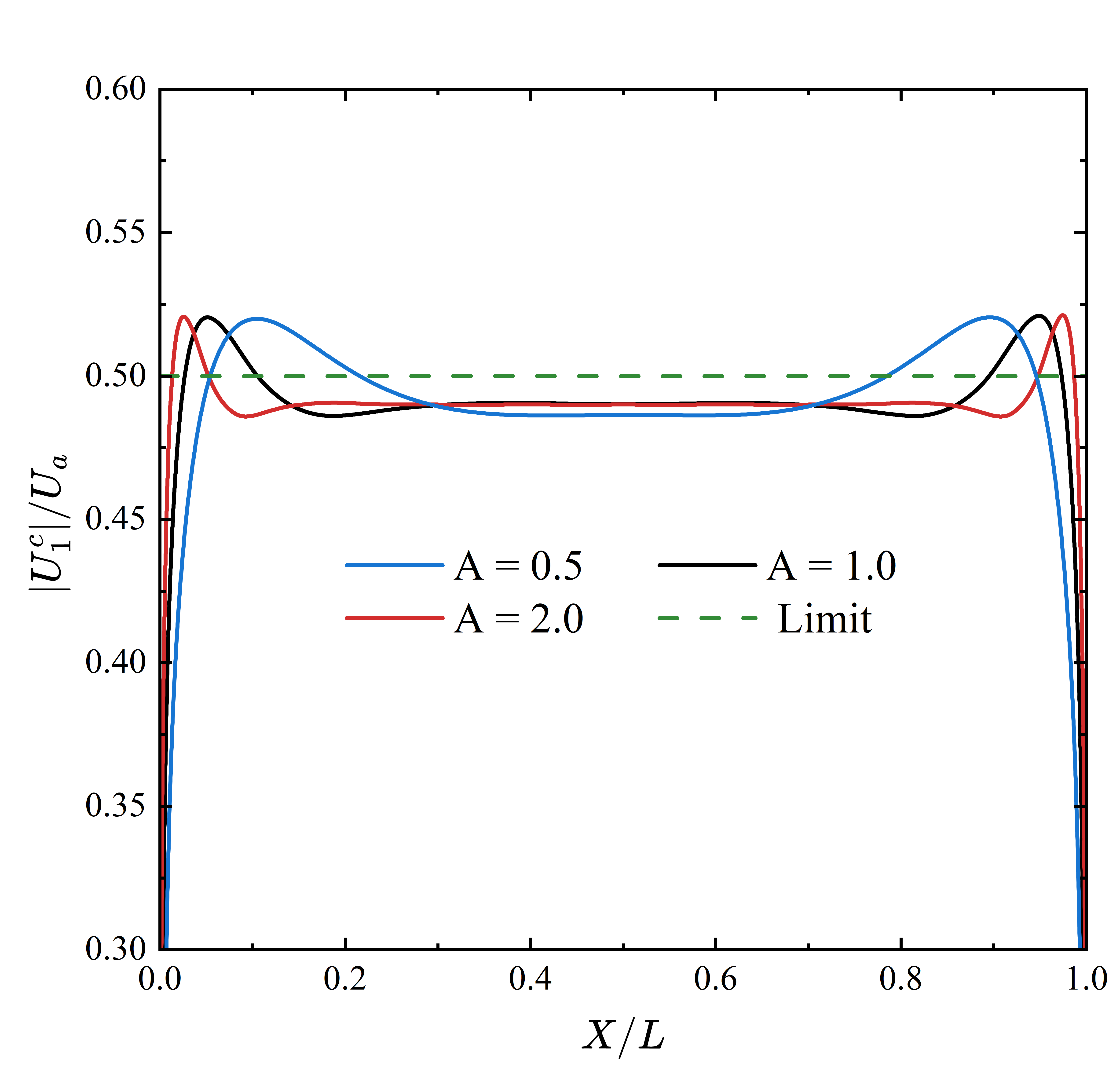}
        \caption{$\widetilde{U}$ along the oscillating lid}
    \end{subfigure}
    \hfill
    \begin{subfigure}{0.48\textwidth}
        \includegraphics[width=\linewidth]{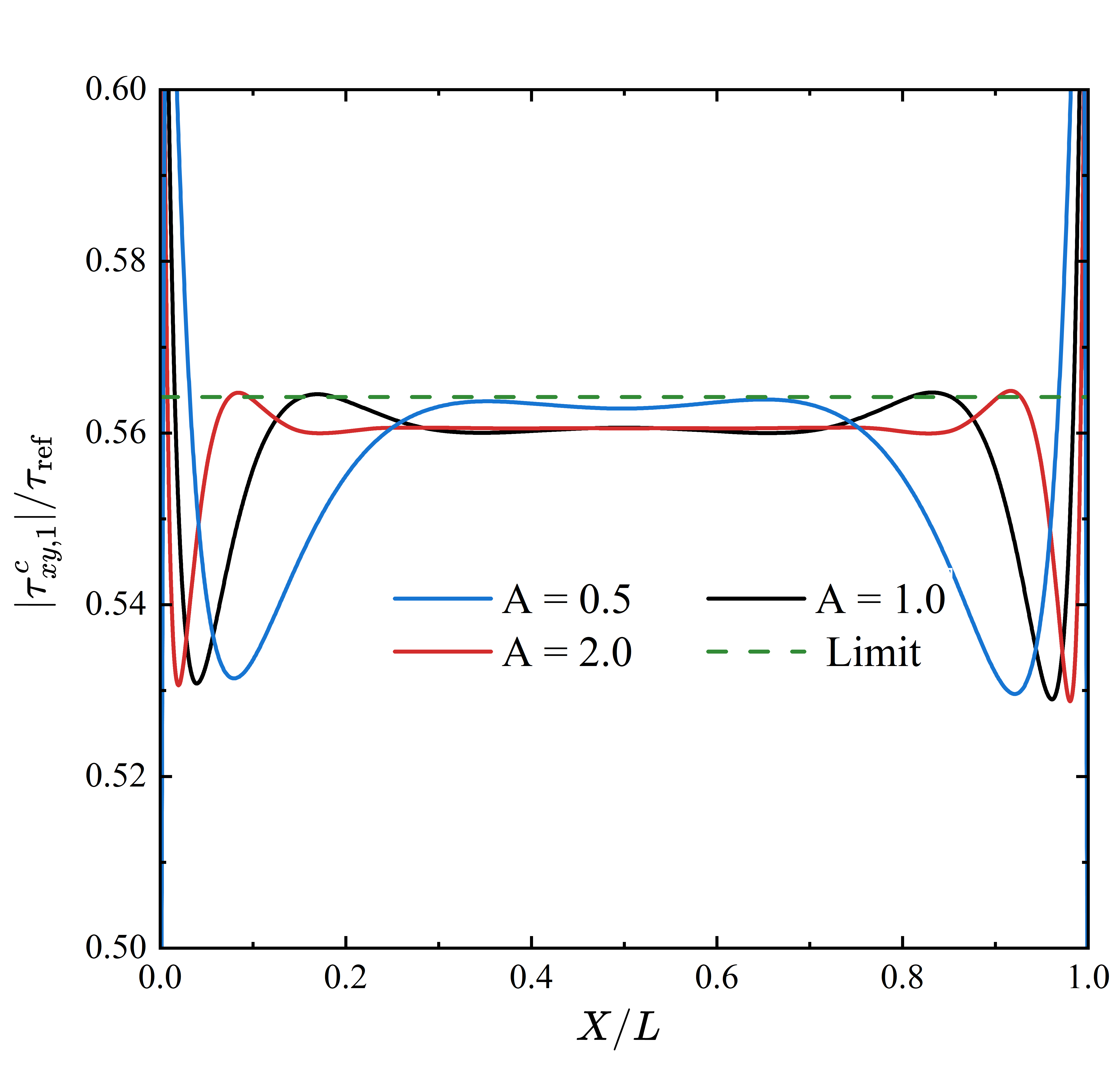}
        \caption{$\widetilde{\tau}_{xy}$ along the oscillating lid}
    \end{subfigure}
    \caption{Distributions of normalized fundamental amplitudes along the lid at $St = 30$ ($Kn = 0.1$) across different aspect ratios. Dashed lines: exact analytical collisionless limits.}
    \label{fig:high_freq_limit}
\end{figure}

\subsubsection{Multiscale regime transition and continuum limit verification}

\begin{figure}[htbp]
    \centering
    \begin{subfigure}{0.48\textwidth}
        \includegraphics[width=\linewidth]{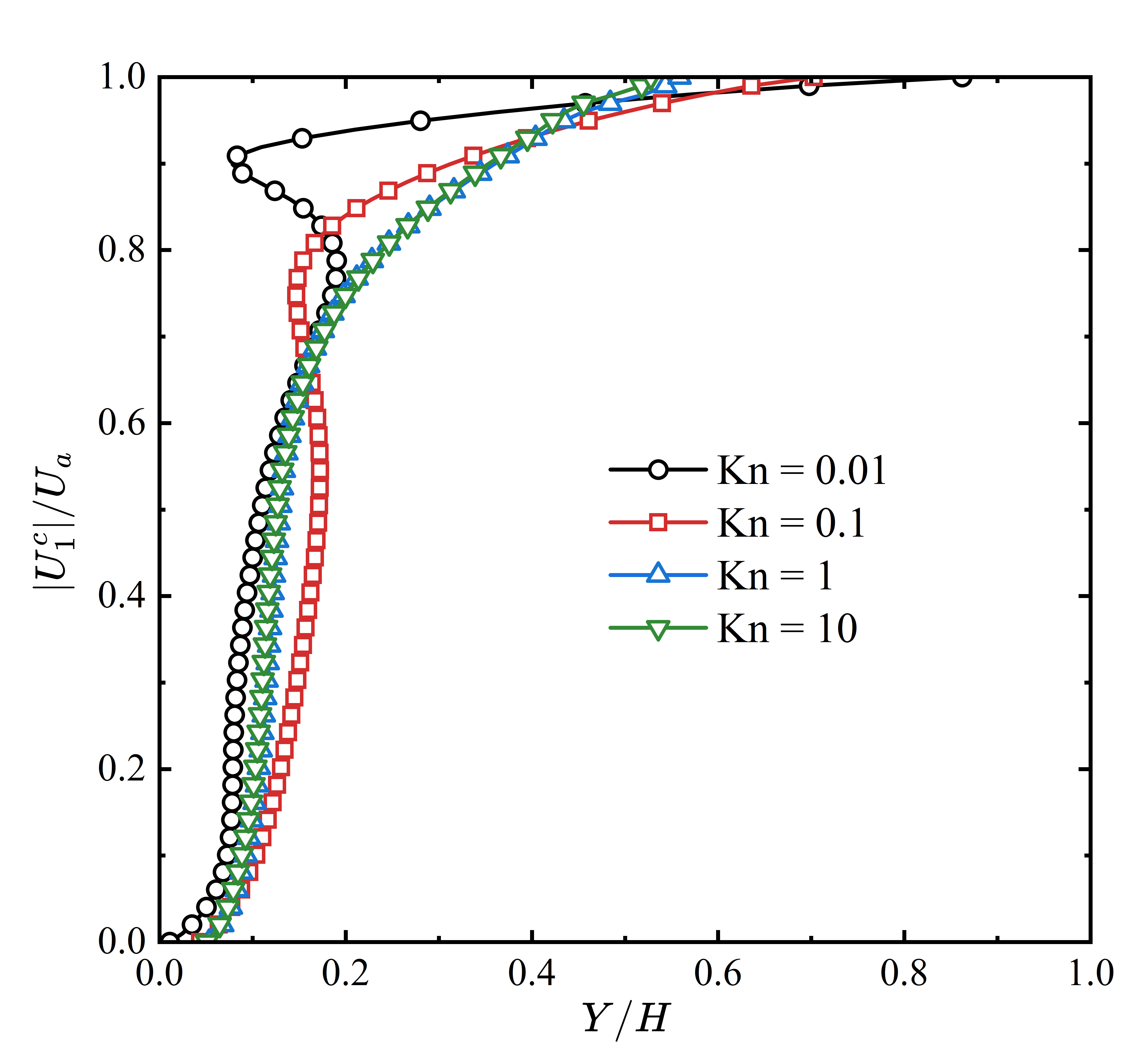}
        \caption{Centerline velocity $|U_1^c|/U_a$ ($x = 0.5L$)}
    \end{subfigure}
    \hfill
    \begin{subfigure}{0.48\textwidth}
        \includegraphics[width=\linewidth]{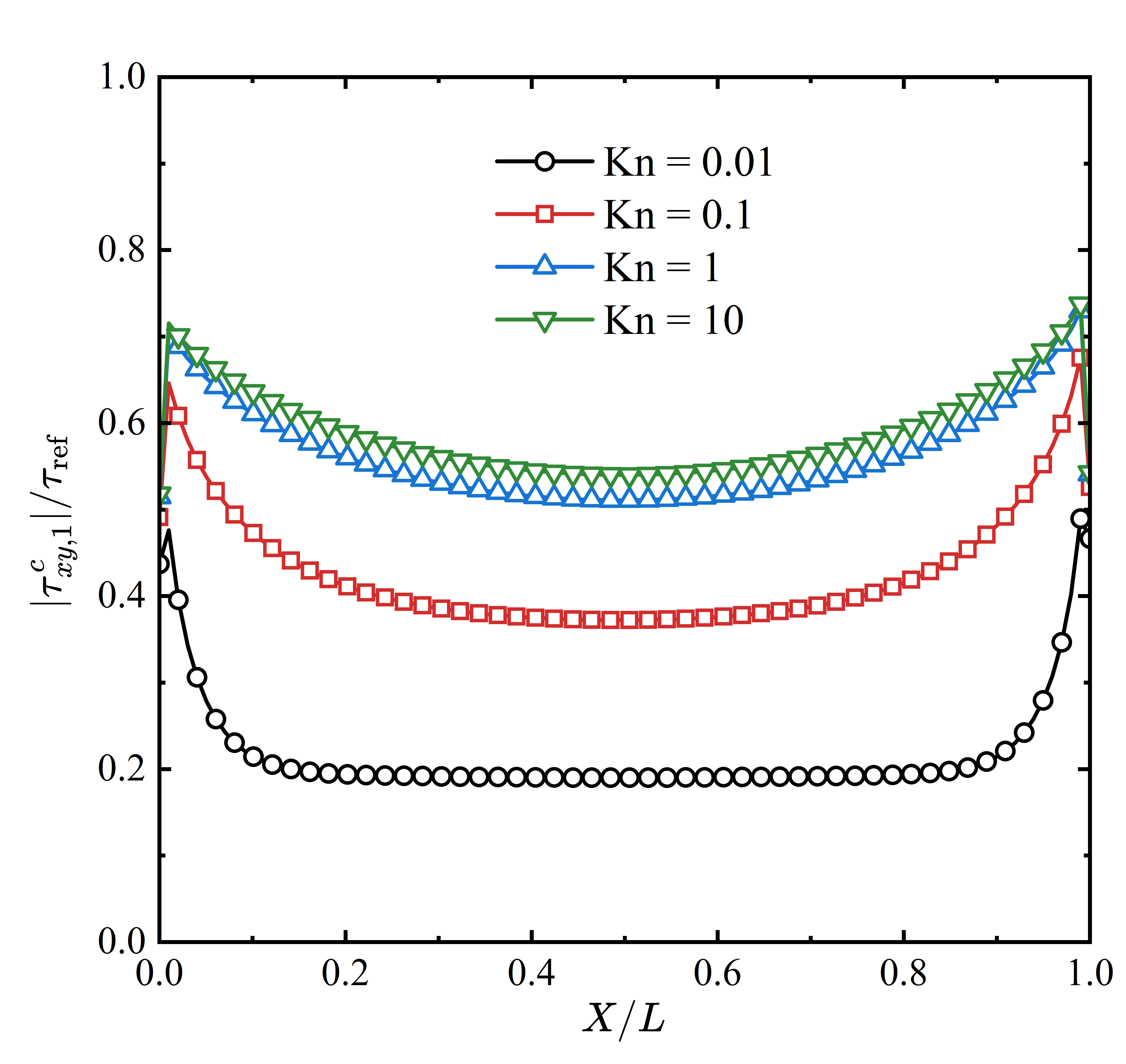}
        \caption{Lid shear stress $|\tau_{xy,1}^c|/\tau_{\text{ref}}$ ($y = H$)}
    \end{subfigure}
    \caption{All-$Kn$ evolution of fundamental harmonic amplitudes at $St = 2$ and $A = 1$ from continuum to rarefied regimes.}
    \label{fig:spectral_profiles_AllKn}
\end{figure}

To examine the multiscale transport properties captured by the coupled interface flux, Fig.~\ref{fig:spectral_profiles_AllKn} displays the fundamental harmonic amplitudes of the centerline velocity and wall shear stress across three orders of magnitude of Knudsen numbers ($Kn = 0.01$ to $10$) at $St = 2$ and $A = 1$. In the near-continuum regime ($Kn = 0.01$), the velocity profile resolves a classical Stokes-layer structure characterized by a spatial velocity reversal near the moving lid ($y/H \approx 0.8 \sim 0.95$) with negligible wall slip, allowing viscous diffusion to penetrate deeply into the cavity. Correspondingly, the wall shear stress exhibits a pronounced U-shaped distribution with sharp corner amplifications caused by two-dimensional wave interactions. As $Kn$ increases, frequent intermolecular collisions diminish, suppressing spatial wave reflections. The velocity profile progressively transitions into a monotonic, shear-dominated distribution accompanied by massive velocity slip, while momentum penetration into the cavity interior is severely restricted. This rarefaction effect is equally evident on the lid, where the wall shear stress flattens into a virtually uniform distribution at $Kn = 10$.

\begin{figure}[htbp]
    \centering
    \begin{subfigure}{0.48\textwidth}
        \includegraphics[width=\linewidth]{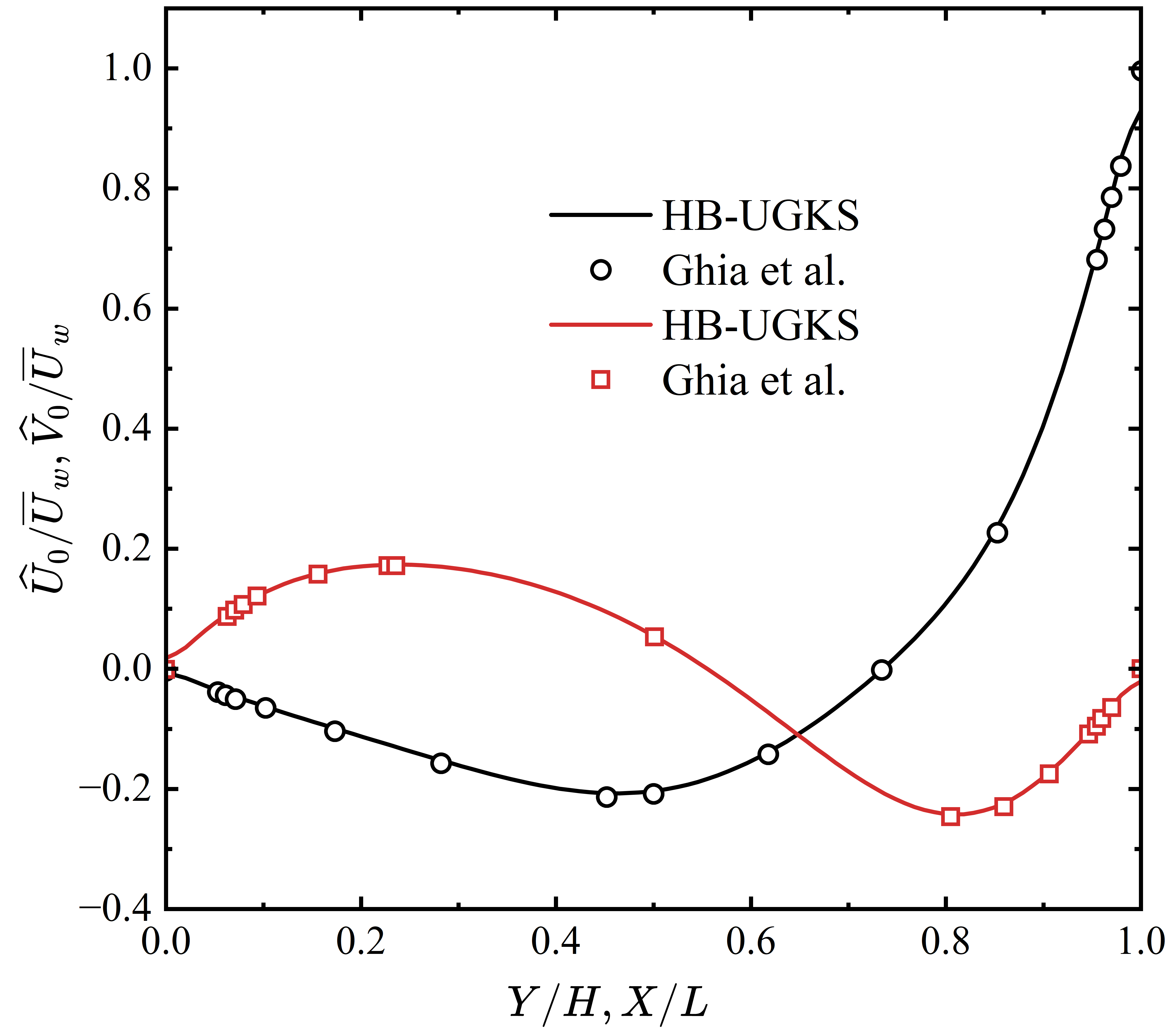}
        \caption{$Re = 100$}
    \end{subfigure}
    \hfill
    \begin{subfigure}{0.48\textwidth}
        \includegraphics[width=\linewidth]{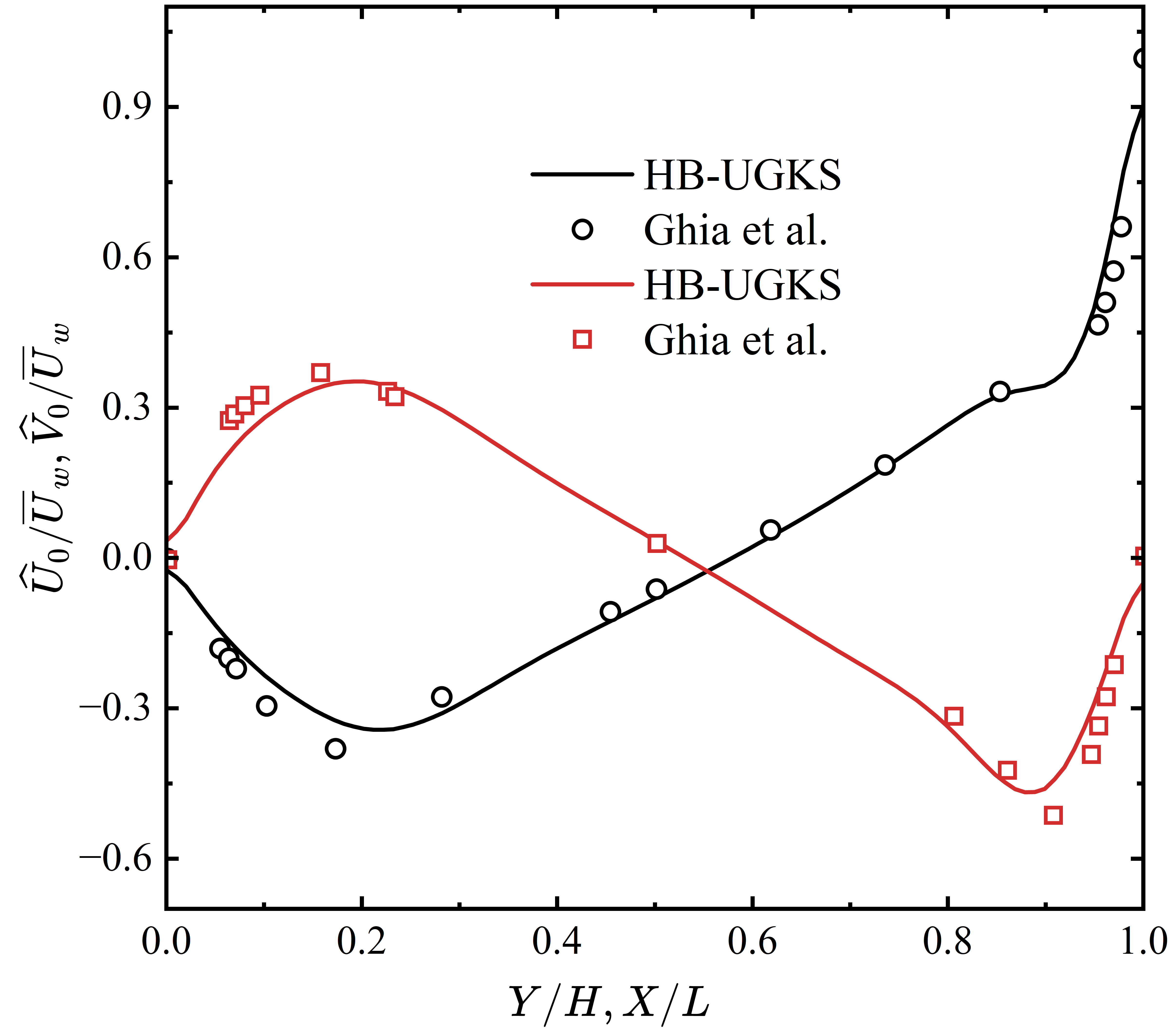}
        \caption{$Re = 1000$}
    \end{subfigure}
    \caption{Continuum limit verification: normalized zeroth-harmonic velocities $\widehat{U}_0/\overline{U}_w$ and $\widehat{V}_0/\overline{U}_w$ along the cavity centerlines compared with benchmark Navier--Stokes data from Ghia et al.~\cite{ghia1982high}.}
    \label{fig:continuum_limit}
\end{figure}

Finally, to demonstrate that the HB-UGKS inherits the multiscale asymptotic-preserving (AP) property of the standard UGKS toward the macroscopic hydrodynamic limit, the cavity is simulated under a mean driving velocity $\overline{U}_w = 0.15$ perturbed by a $1\%$ oscillation amplitude ($U_a = 0.0015$). In this weakly perturbed regime, the unsteady forcing does not distort the background field, allowing the zeroth-order harmonic component $\widehat{\mathbf{W}}_0$ to represent the steady-state hydrodynamic solution. Figure~\ref{fig:continuum_limit} compares the computed time-averaged centerline velocities at $Re = 100$ and $Re = 1000$ with the classical benchmark data of Ghia et al.~\cite{ghia1982high}. The excellent agreement for both Reynolds numbers confirms that the coupled multiscale flux seamlessly recovers the Navier--Stokes solutions in the continuum limit without grid-size restrictions associated with the mean free path.

\subsubsection{Computational efficiency of the HB-UGKS}
\label{sec:shear_efficiency}

The harmonic balance framework is devised not only to capture the periodic solution accurately but also to offer significant potential for reducing computational time compared to conventional physical-time marching algorithms. To ensure a rigorous and consistent assessment, both the HB-UGKS and the conventional explicit time-domain (TD) UGKS are implemented within the same codebase and executed on an identical high-performance computing platform. They share the same spatial meshes, velocity-space discretizations, interface reconstruction schemes, boundary conditions, and parallelization configurations. The distinction lies in the temporal treatment: while the TD method explicitly advances physical time step-by-step, the HB-UGKS converts the physical-time derivative into a spectral operator, recasting the unsteady problem into a coupled system of steady-state equations converged via pseudo-time iterations.

\begin{figure}[htbp]
    \centering
    \includegraphics[width=0.5\linewidth]{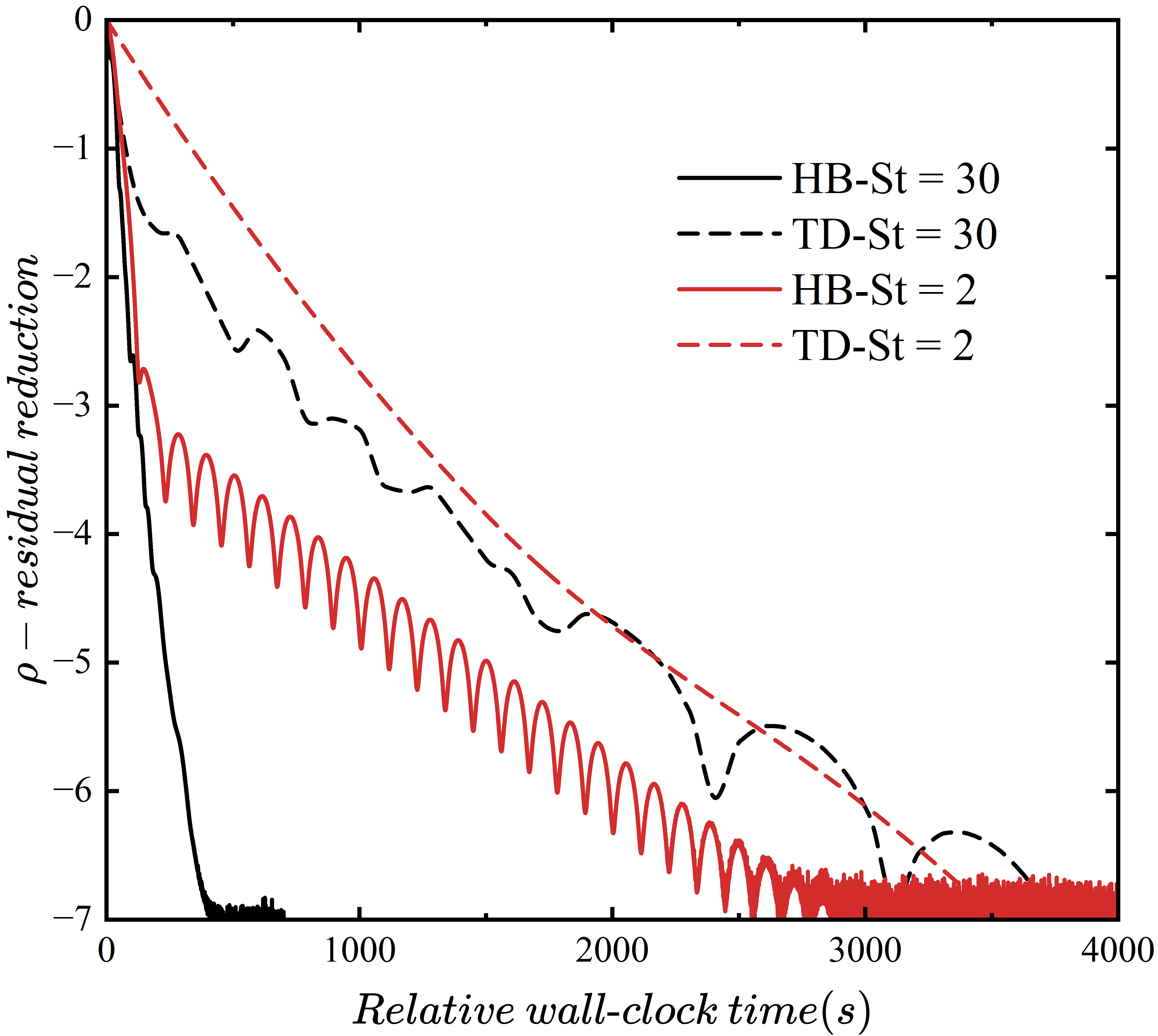}
    \caption{Convergence comparison between the HB-UGKS and explicit time-domain UGKS for the shear-driven cavity under the unified benchmark definition of Eq.~(\ref{eq:benchmark_residuals}).}
    \label{fig:wu_speedup}
\end{figure}

For a conservative variable $W^{(p)} \in \{\rho,\rho U,\rho V,\rho E\}$, the residuals used to evaluate convergence toward the periodic state are defined as:
\begin{equation}
\begin{aligned}
\mathcal{R}_{\mathrm{HB}}^{(p)}(k)&=
\left[
\frac{1}{N_c}
\sum_{i=1}^{N_c}
\left(
W_{i,0}^{(p),k+1}-W_{i,0}^{(p),k}
\right)^2
\right]^{1/2},\\
\mathcal{R}_{\mathrm{TD}}^{(p)}(n)&=
\left[
\frac{1}{N_c}
\sum_{i=1}^{N_c}
\left(
W_i^{(p)}(nT)-W_i^{(p)}((n-1)T)
\right)^2
\right]^{1/2},
\end{aligned}
\label{eq:benchmark_residuals}
\end{equation}
where $N_c$ is the total cell count. For the HB-UGKS, the residual measures the pseudo-step update at the initial temporal collocation point ($t_0 = 0$). For the TD solver, the residual tracks the period-to-period change at the corresponding phase across consecutive cycles. To ensure an unambiguous visual comparison, the convergence histories are presented in Fig.~\ref{fig:wu_speedup} using the shifted density-residual reduction, $\Delta r_\rho = \log_{10}\mathcal{R}_{\rho}-\log_{10}\mathcal{R}_{\rho,1}$, against the shifted wall-clock time, $\Delta t_{\mathrm{wall}} = t_{\mathrm{wall}}-t_{\mathrm{wall},1}$.

Figure~\ref{fig:wu_speedup} compares the wall-clock time required for the density residual to reach convergence at two representative frequencies. At the lower frequency of $St = 2$, the HB-UGKS achieves a moderate speedup factor of approximately $1.4\times$. In this regime, the forcing period is relatively long, and the physical relaxation time required to wash out startup transients spans only a few driving cycles; hence, the explicit time-marching cost remains manageable.

At $St = 30$, however, the speedup surges dramatically to $9.2\times$. This stark contrast highlights the inherent weakness of physical-time marching for high-frequency periodic flows. While the physical settling time required to reach the stable periodic state remains roughly constant, the duration of a single excitation cycle shrinks substantially at high $St$. Consequently, the explicit TD solver must march through dozens of physical periods (often exceeding 50 cycles) simply to accumulate enough physical time to stabilize, rendering it computationally prohibitive. By contrast, the HB formulation entirely bypasses this cycle-by-cycle settling process by directly solving for the invariant limit-cycle state, thereby demonstrating superior computational efficiency for high-frequency oscillatory non-equilibrium flows.

\FloatBarrier
\subsection{Oscillatory shear-driven flow in an eccentric annulus}
\label{sec:eccentric_annulus}

The second benchmark extends the small-amplitude mechanical oscillation to curved boundaries with a non-uniform clearance. The obtained flow responses are validated against reference solutions to ensure numerical accuracy, while the computational efficiency of the HB-UGKS is systematically evaluated through direct comparisons with the explicit TD solver.

\subsubsection{Numerical configuration and setups}

\begin{figure}[htbp]
    \centering
    \includegraphics[width=0.5\linewidth]{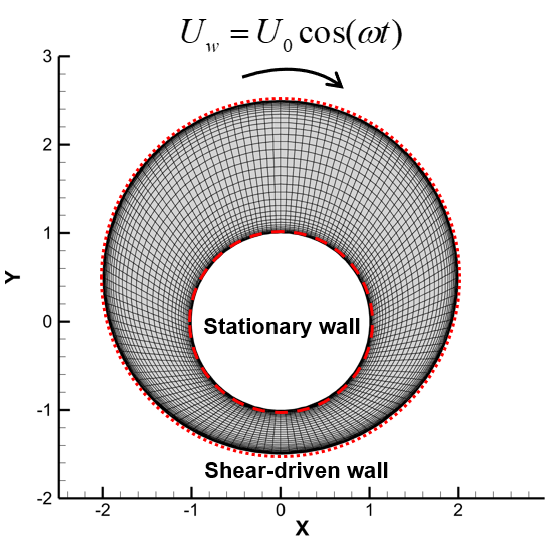}
    \caption{Geometry and spatial mesh of the eccentric-annulus benchmark. The inner cylinder is stationary, while the outer cylinder undergoes harmonic tangential motion with speed $U_w=U_0\cos(\omega t)$. Coordinates are normalized by the inner-cylinder radius $L$.}
    \label{fig:eccentric_mesh}
\end{figure}

The computational configuration and spatial discretization are illustrated in Fig.~\ref{fig:eccentric_mesh}. The flow domain is bounded by two eccentric circular cylinders, with the inner cylinder of radius $R_i = L$ centered at the origin $(0, 0)$ and the outer cylinder of radius $R_o = 2L$ centered at $(0, 0.5L)$. This arrangement yields a non-dimensional eccentricity of $e/L = 0.5$, resulting in non-uniform vertical clearances of $1.5L$ above the inner cylinder and $0.5L$ below it. Both cylinder walls are maintained at an isothermal temperature $T_0$ under fully diffuse reflection conditions. While the inner cylinder remains stationary, the outer wall is subjected to a harmonically oscillating tangential velocity:
\begin{equation}
\mathbf{U}_w(\mathbf{x}, t) = U_0 \cos(\omega t) \, \mathbf{t}_w(\mathbf{x}),
\label{eq:eccentric_wall}
\end{equation}
where $\mathbf{t}_w(\mathbf{x})$ denotes the unit tangent vector pointing in the positive $x$-direction at the top of the outer cylinder. The driving amplitude is specified as $U_0 = \epsilon v_m$ with $\epsilon = 10^{-3}$, where $v_m = \sqrt{2RT_0}$ is the most probable molecular speed. Because the oscillatory motion is purely tangential along the circular profile, the physical boundary remains geometrically invariant in time, naturally avoiding the complexity of moving or deforming meshes.

To facilitate direct comparison with literature benchmarks, the flow regime is characterized by the rarefaction parameter $\delta$ and the dimensionless frequency parameter $S$:
\begin{equation}
\delta = \frac{p_0 L}{\mu_0 v_m},
\label{eq:def_delta}
\end{equation}
\begin{equation}
S = \frac{\omega L}{v_m},
\label{eq:def_S}
\end{equation}
where the inner radius $L$ serves as the characteristic length scale, $p_0$ is the equilibrium pressure, and $\mu_0$ is the dynamic viscosity at $T_0$. The ratio of these two parameters defines the temporal Knudsen number:
\begin{equation}
Kn_t = \frac{S}{\delta} = \omega \tau_0,
\label{eq:def_Knt}
\end{equation}
which characterizes the ratio of the molecular collision time $\tau_0$ to the which characterizes the ratio of the molecular collision time
$\tau_0$ to the oscillation time scale $1/\omega$.

The physical domain is discretized using a structured grid comprising $132 \times 46$ cells in the circumferential and radial directions, respectively, with cells clustered toward both solid walls to adequately resolve near-wall kinetic layers. The velocity space is truncated to $[-6v_m, 6v_m]^2$ and discretized by a non-uniform $32 \times 32$ Cartesian grid. Consistent with the linear response observed in the small-amplitude regime, the fundamental truncation order is set to $N_\text{H} = 1$ . A systematic parametric study is conducted across three rarefaction regimes ($\delta = 1, 10, 100$) over a wide frequency spectrum of $S \in [0, 30]$, where the limiting case of $S = 0$ corresponds to the steady Couette-type flow.

\subsubsection{Fourier-field verification against the time-domain solution}

To verify the accuracy of the HB formulation and demonstrate that retaining a fundamental harmonic ($N_\text{H} = 1$) is sufficient in the weakly oscillating regime, the computed fundamental Fourier fields are compared directly against those from the TD solver. For the HB-UGKS, the two-sided Fourier coefficients $\widehat{q}_1$ are obtained from Eq.~(\ref{eq:dft_general}), and the corresponding one-sided complex harmonic amplitudes
$q_1^c=2\widehat{q}_1$ are used for the following comparisons.

\begin{figure}[h!]
    \centering
    \begin{subfigure}{0.48\textwidth}
        \centering
        \includegraphics[width=\linewidth]{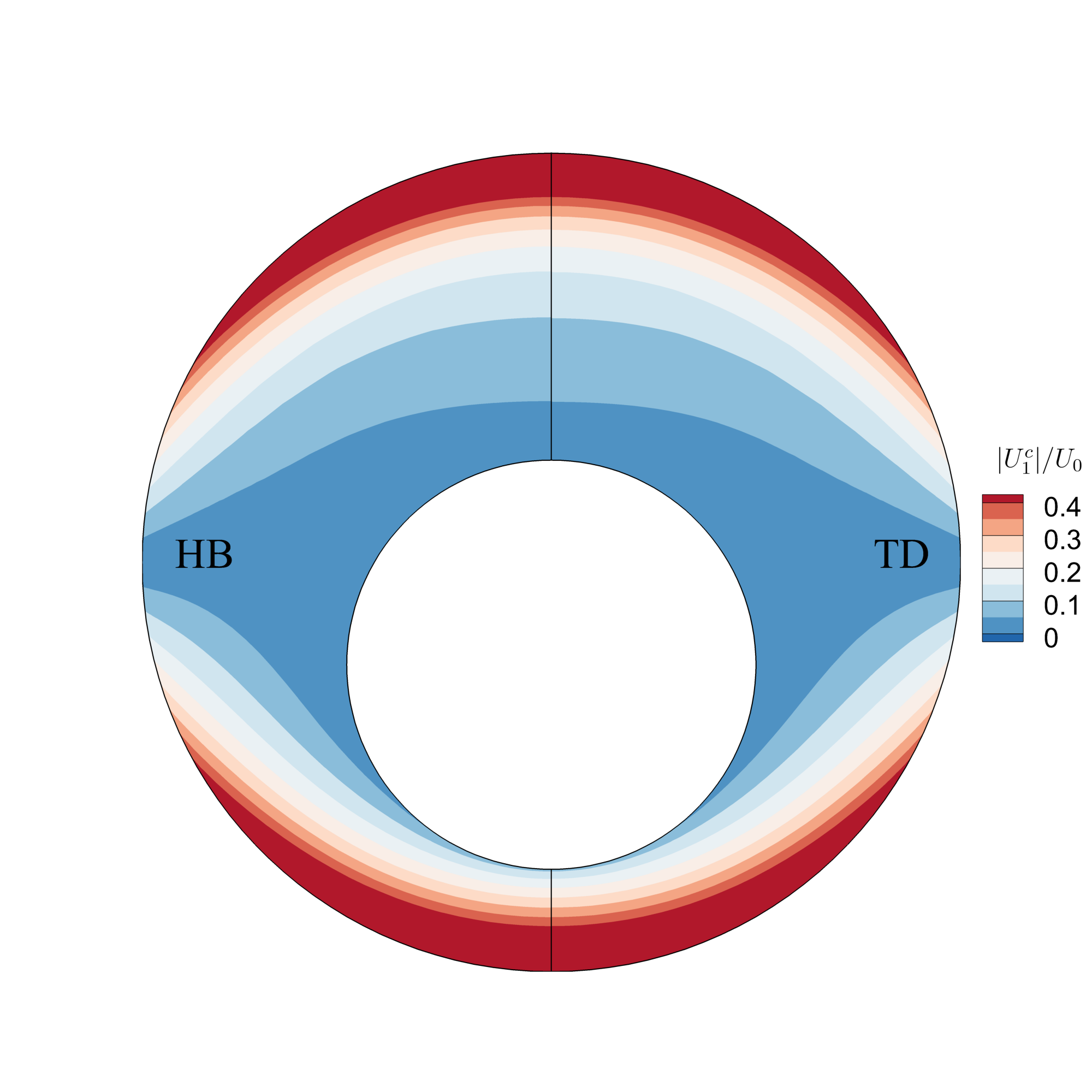}
        \caption{Horizontal velocity, $S=1$.}
        \label{fig:eccentric_u_s1}
    \end{subfigure}
    \hfill
    \begin{subfigure}{0.48\textwidth}
        \centering
        \includegraphics[width=\linewidth]{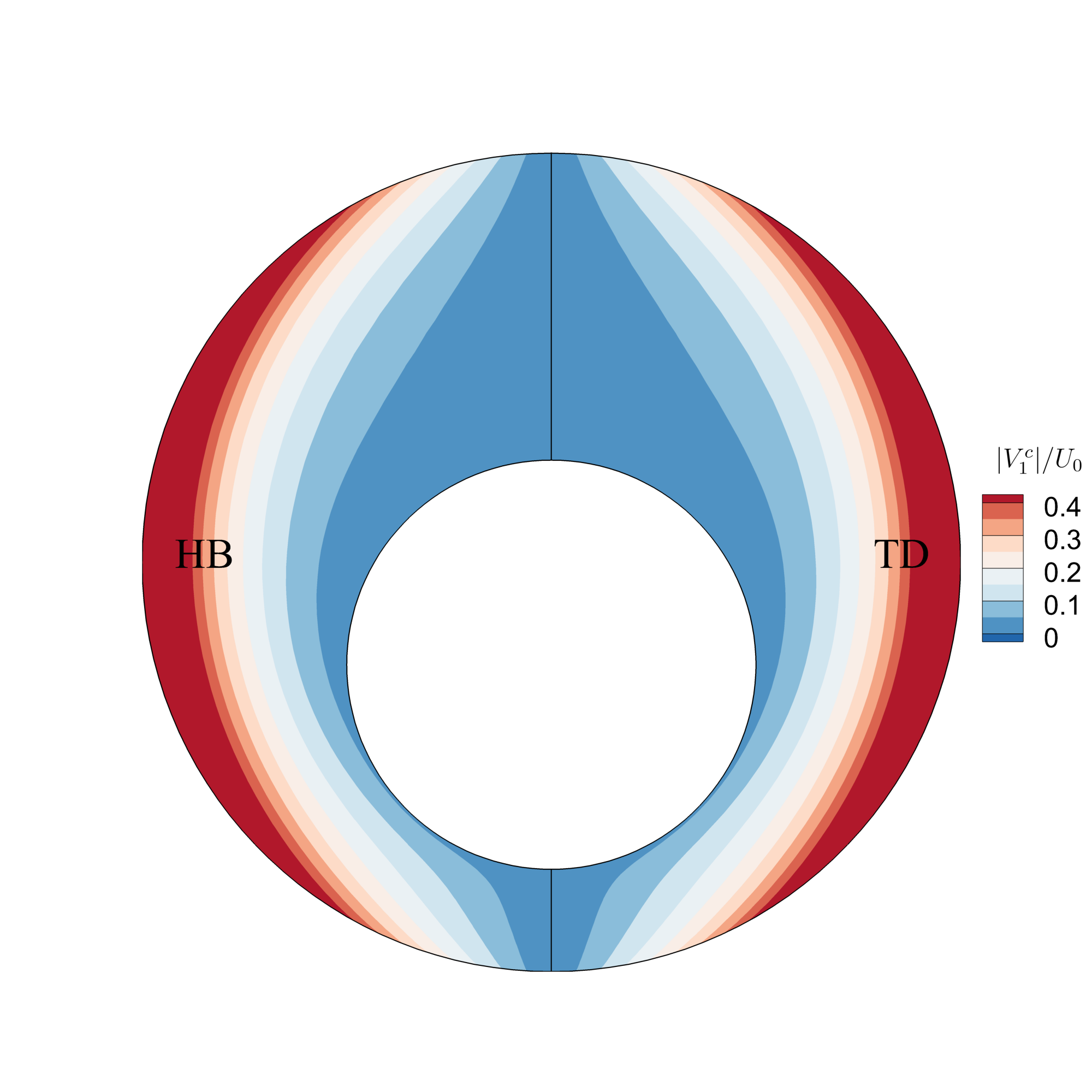}
        \caption{Vertical velocity, $S=1$.}
        \label{fig:eccentric_v_s1}
    \end{subfigure}
    \par\medskip
    \begin{subfigure}{0.48\textwidth}
        \centering
        \includegraphics[width=\linewidth]{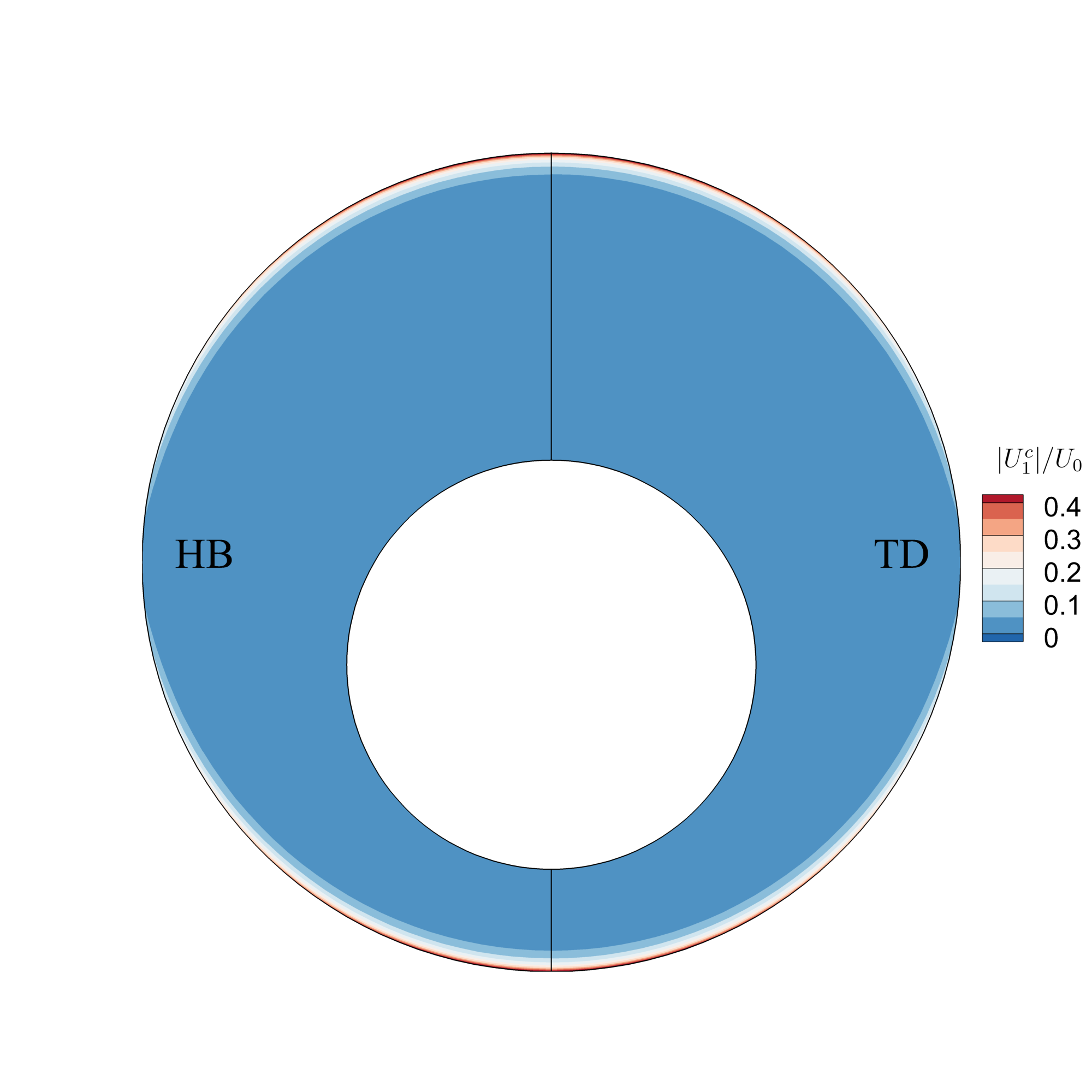}
        \caption{Horizontal velocity, $S=30$.}
        \label{fig:eccentric_u_s30}
    \end{subfigure}
    \hfill
    \begin{subfigure}{0.48\textwidth}
        \centering
        \includegraphics[width=\linewidth]{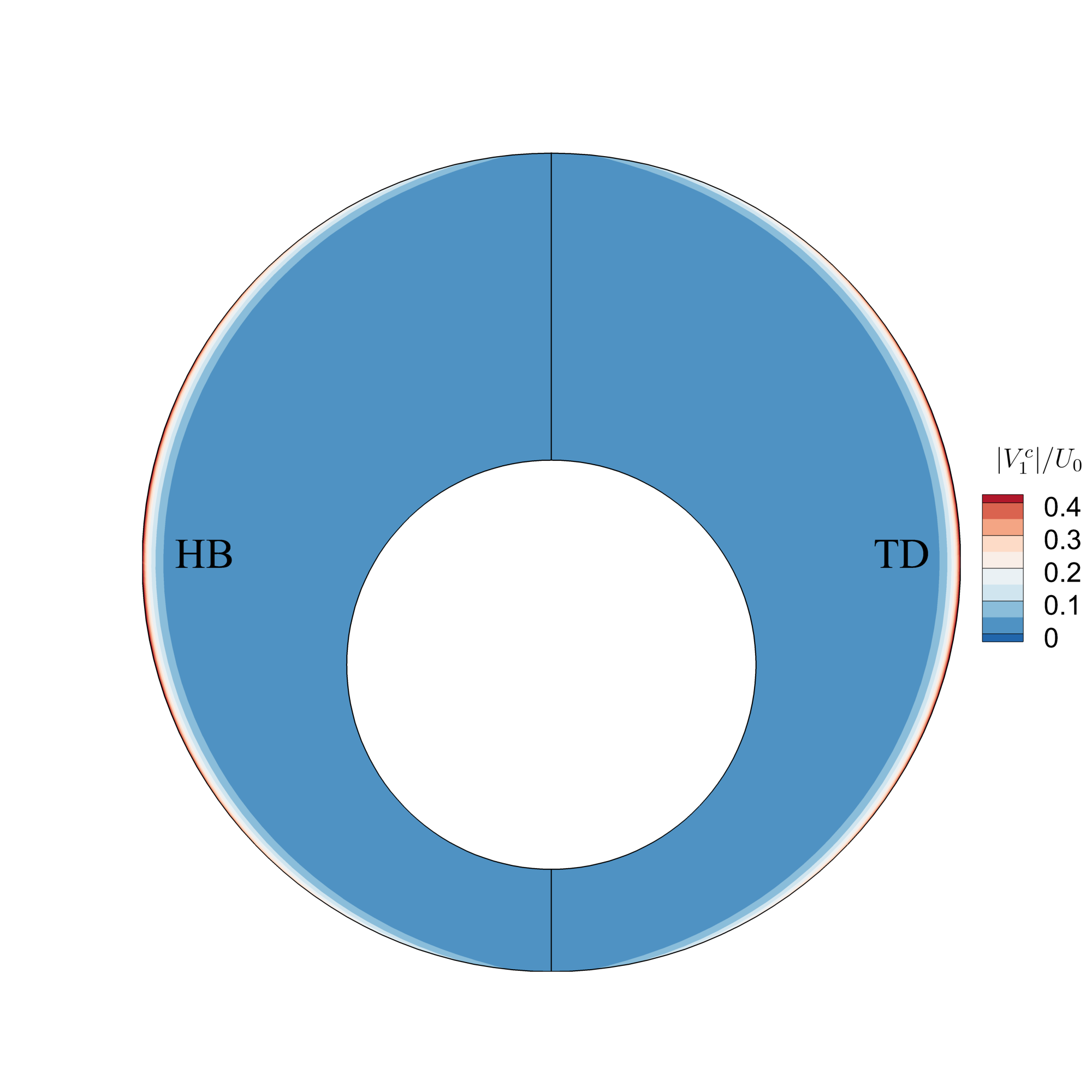}
        \caption{Vertical velocity, $S=30$.}
        \label{fig:eccentric_v_s30}
    \end{subfigure}
    \caption{Comparison of normalized fundamental velocity amplitudes between HB-UGKS (left half) and time-domain UGKS (right half) in the eccentric annulus at $\delta=10$.}
    \label{fig:eccentric_fourier}
\end{figure}

Figure~\ref{fig:eccentric_fourier} compares the normalized fundamental velocity magnitudes,
$|U_1^c|/U_0$ and $|V_1^c|/U_0$, between the HB and TD solutions at $\delta = 10$ for two contrasting frequencies: $S = 1$ and $S = 30$. In each panel, the left half presents the HB result and the right half displays the TD solution. The contour lines connect seamlessly across the vertical interface between the two halves, demonstrating excellent agreement and confirming that the first harmonic completely governs the dynamic response.

The spatial profiles reveal distinctly different flow physics governed by the driving frequency. In the low-frequency regime ($S = 1$), viscous diffusion operates on a timescale comparable to the oscillation period, allowing shear momentum to penetrate deeply across the annular clearance. Consequently, the oscillatory disturbances spread throughout the entire eccentric domain and interact directly with the stationary inner cylinder. The spatial distribution reflects the geometric orientation of the boundary motion: $|U_1^c|$ reaches its maximum near the top and bottom of the outer cylinder where the wall velocity is predominantly horizontal (Fig.~\ref{fig:eccentric_u_s1}), whereas $|V_1^c|$ concentrates along the lateral sides where the motion is purely vertical (Fig.~\ref{fig:eccentric_v_s1}). 

Conversely, in the high-frequency regime ($S = 30$), the forcing period becomes substantially shorter than the viscous diffusion time across the gap. The flow transitions into a classic Stokes boundary-layer regime, where dynamic perturbations are strictly confined to a thin layer immediately adjacent to the moving outer wall (Figs.~\ref{fig:eccentric_u_s30} and \ref{fig:eccentric_v_s30}). The core region and the inner cylinder remain virtually unperturbed, with velocity amplitudes decaying rapidly toward zero away from the driving surface. The seamless matching between the HB and TD solutions across both frequency regimes confirms that the single-harmonic HB-UGKS accurately captures the multiscale transport mechanisms under complex curved geometries.

\subsubsection{Wall-load frequency response}

The frequency response of the integrated tangential wall load on the outer cylinder is examined across three representative flow regimes: $\delta = 1$ (rarefied), $\delta = 10$ (transitional), and $\delta = 100$ (near-continuum). This macroscopic quantity simultaneously captures the amplitude and phase characteristics of the dynamic shear traction.

The dimensionless first-harmonic tangential wall load $\gamma$ is defined by integrating the complex wall shear traction along the outer boundary $\Gamma_o$:
\begin{equation}
\gamma =
\frac{1}{p_0\epsilon L}
\int_{\Gamma_o}\tau_{w,1}^{c}(\mathbf{x})\,\mathrm{d}s,
\end{equation}
where $\tau_{w,1}^{c}$ denotes the one-sided complex amplitude of the first-harmonic tangential shear stress exerted by the gas on the wall, and the factor $p_0 \epsilon L$ provides the characteristic force scale. The dynamic response is fully quantified by the load amplitude $\mathcal{A}_\gamma$ and phase angle $\phi_\gamma$:
\begin{equation}
    \mathcal{A}_\gamma = |\gamma|, \qquad \phi_\gamma = \arg(\gamma).
    \label{eq:load_amp_phase}
\end{equation}
Here, $\phi_\gamma$ is defined with respect to the prescribed cosine wall velocity in Eq.~(\ref{eq:eccentric_wall}). Under the Fourier convention $q(t)=\Re(q_1^c e^{i\omega t})$, the imposed wall velocity has zero phase. Therefore, $\phi_\gamma=\arg(\gamma)$ measures the phase of the gas-on-wall tangential load relative to the wall velocity. A positive $\phi_\gamma$ indicates that the load leads the
wall velocity, while a negative $\phi_\gamma$ indicates a lag. In the present work, the tangential traction and the integrated load are defined as the force exerted by the gas on the wall, with the positive direction taken along the local tangential unit vector
$\mathbf{t}_w$.

\begin{figure}[htbp]
    \centering
    \begin{subfigure}{0.48\textwidth}
        \centering
        \includegraphics[width=\linewidth]{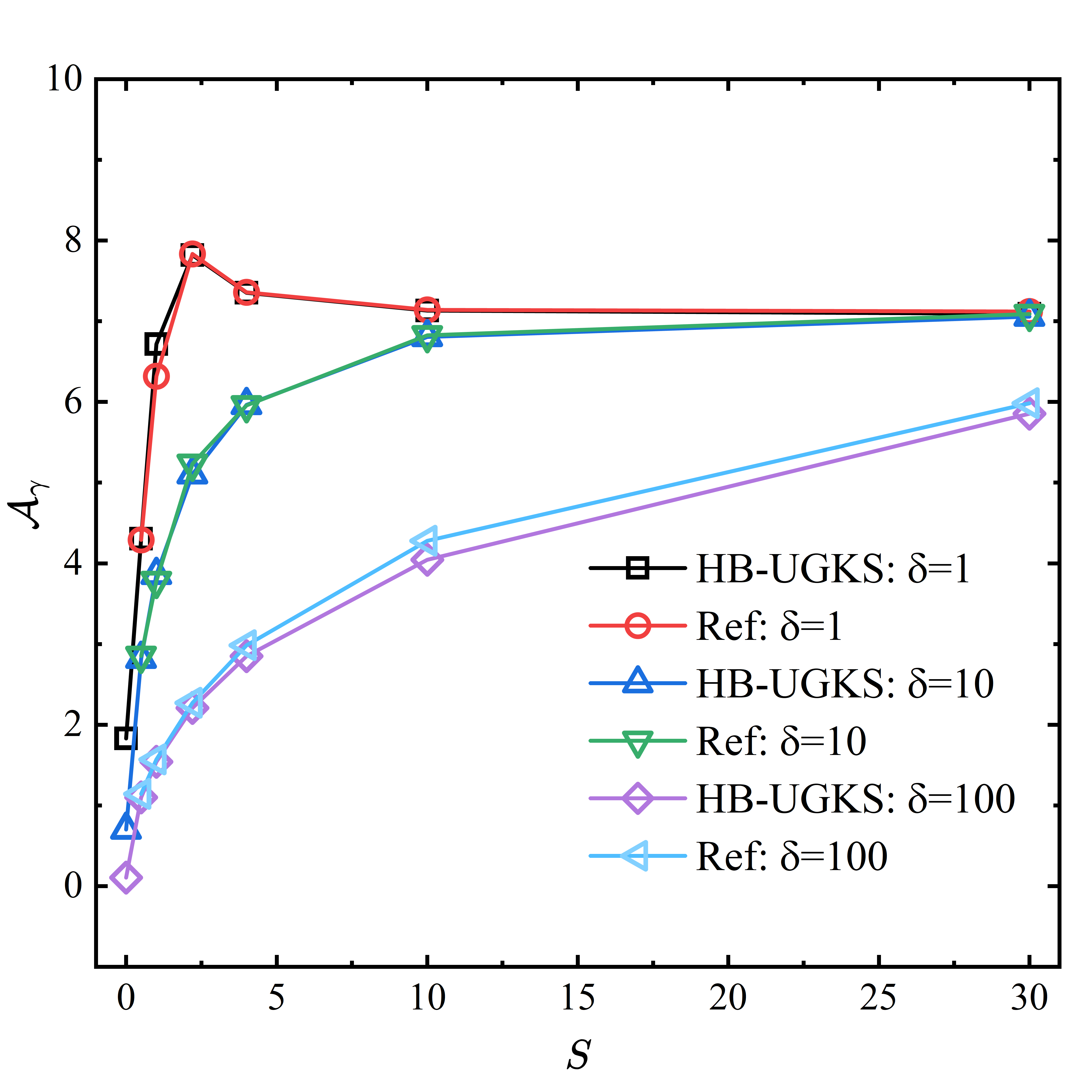}
        \caption{Load amplitude $\mathcal{A}_\gamma$.}
        \label{fig:eccentric_load_amplitude}
    \end{subfigure}
    \hfill
    \begin{subfigure}{0.48\textwidth}
        \centering
        \includegraphics[width=\linewidth]{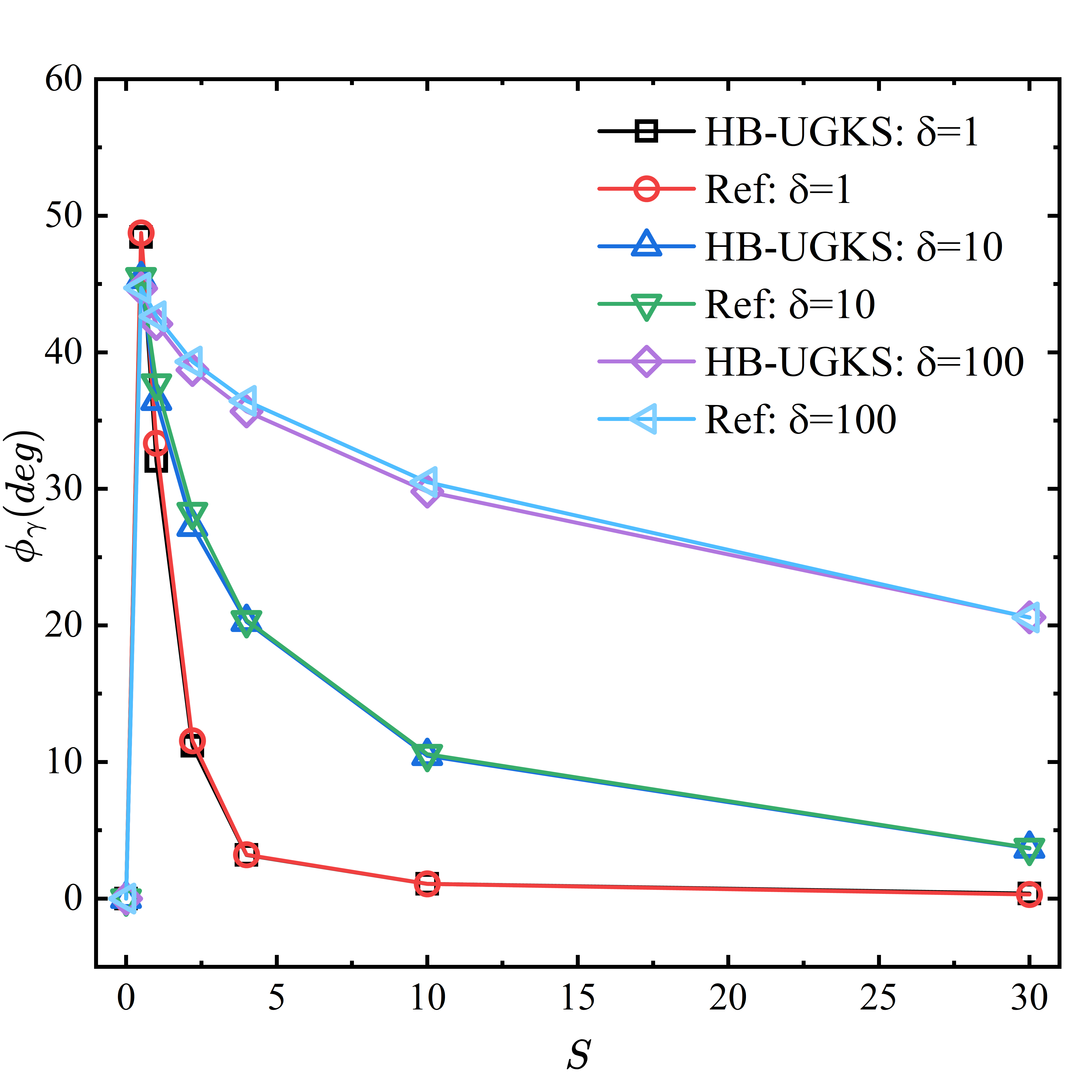}
        \caption{Load phase $\phi_\gamma$ (degrees).}
        \label{fig:eccentric_load_phase}
    \end{subfigure}
    \caption{Frequency response of the integrated tangential wall load on the outer cylinder for $\delta = 1, 10, 100$. Solid curves with symbols represent the present HB-UGKS results, while open markers denote reference data from Li and Wu~\cite{li2026frequency}.}
    \label{fig:eccentric_response}
\end{figure}

Figure~\ref{fig:eccentric_response} presents the variation of $\mathcal{A}_\gamma$ and $\phi_\gamma$ over a wide frequency spectrum spanning from the quasi-steady limit up to $S = 30$. The load amplitude exhibits qualitatively distinct dependencies on the oscillation frequency across different Knudsen regimes. In the rarefied regime ($\delta = 1$), where intermolecular collisions are infrequent, the load amplitude exhibits a slight peak at moderate frequencies before asymptotically leveling off toward the collisionless ballistic limit at high $S$, where direct wall--particle interactions dominate over collective viscous shearing. Conversely, in the transitional and near-continuum regimes ($\delta = 10$ and $100$), the load magnitude increases monotonically with frequency. As $S$ increases, the oscillatory Stokes boundary layer contracts progressively toward the outer boundary, steepening the near-wall velocity gradients and consequently elevating the integrated shear traction.

As shown in Fig.~\ref{fig:eccentric_load_phase}, the phase response exhibits a distinct non-monotonic dependence on the oscillation frequency across all three rarefaction regimes. Starting from the quasi-steady limit where the phase shift remains negligible, the phase angle rises to a positive peak at intermediate values of $S$ and gradually decays toward an asymptotic high-frequency plateau. Although the peak location and subsequent decay rate are quantitatively governed by the rarefaction parameter $\delta$, the HB-UGKS captures this non-monotonic trend in close agreement with the reference solutions across all regimes.

\subsubsection{Computational efficiency and convergence}
\label{sec:eccentric_efficiency}

\begin{figure}[htbp]
    \centering
    \includegraphics[width=0.5\linewidth]{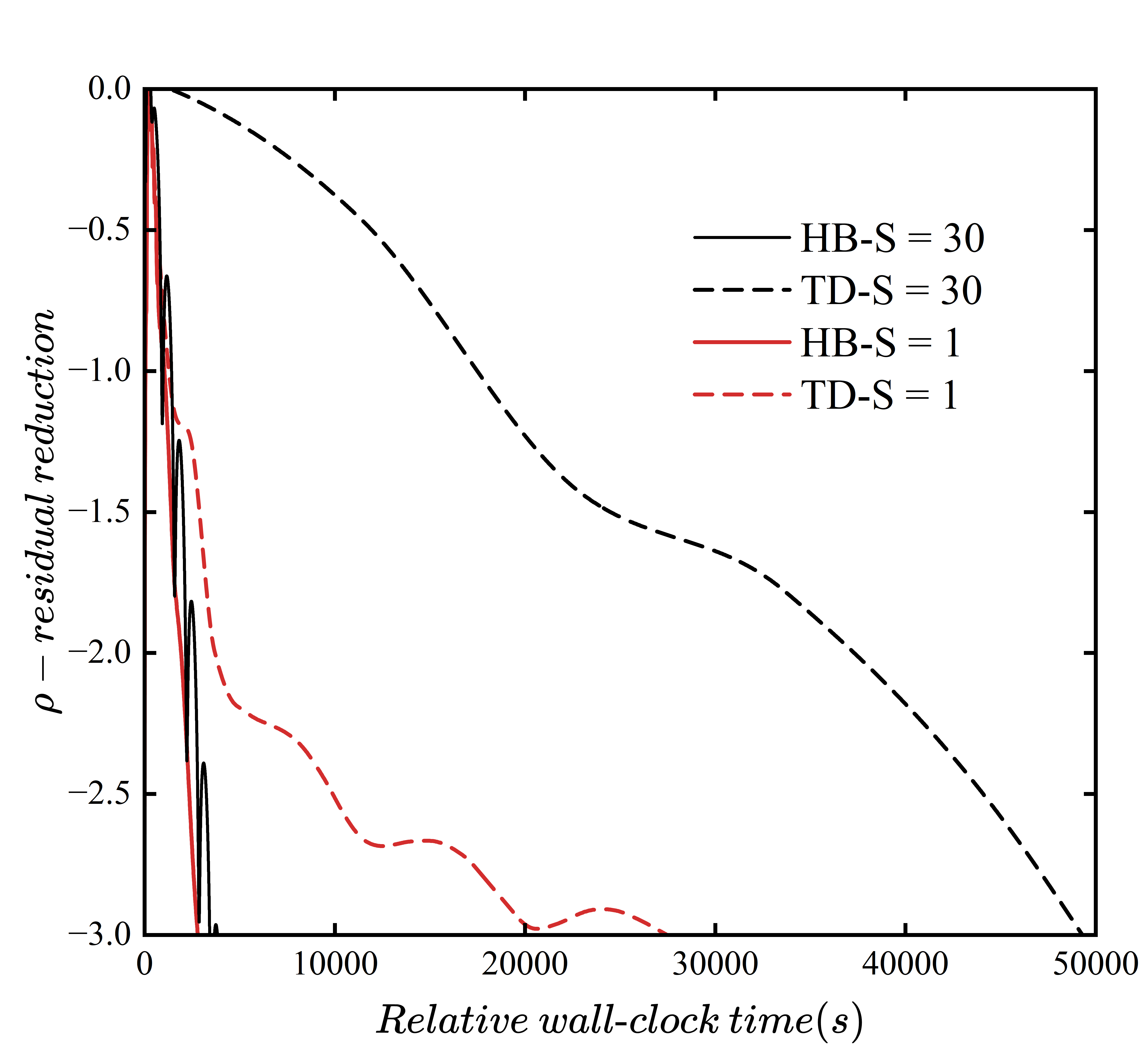}
    \caption{Convergence histories of macroscopic density residual versus wall-clock time for the eccentric annulus at $\delta=10$ with $S=1$ and $S=30$. Solid curves denote the present HB-UGKS, and dashed curves represent the conventional time-domain solver.}
    \label{fig:eccentric_speedup}
\end{figure}

To demonstrate the computational advantage of the harmonic balance formulation, the convergence history of the HB-UGKS is evaluated against the TD solver for the eccentric annulus at $\delta = 10$. Figure~\ref{fig:eccentric_speedup} displays the reduction of the macroscopic density residual as a function of wall-clock time for both low- ($S=1$) and high-frequency ($S=30$) oscillations, recorded on the same computing platform.

As shown in Fig.~\ref{fig:eccentric_speedup}, the HB-UGKS exhibits superior computational efficiency across both frequencies. For both $S=1$ and $S=30$, the pseudo-time residual in the HB method drops by three orders of magnitude within approximately $3.5 \times 10^3$~s. Remarkably, the convergence rate of the HB solver shows little sensitivity to the oscillation frequency, since it directly seeks the periodic steady state by marching the coupled harmonic equations in pseudo-time, completely bypassing the physical transient phase.

In contrast, the convergence of the conventional TD solver is severely hindered by the physical time required to wash out initial transients and attain a strictly periodic state. At $S=1$, the TD simulation takes approximately $2.1 \times 10^4$~s to achieve the same three-order residual drop, yielding an acceleration factor of about $6$ for the HB method. At the higher frequency ($S=30$), the disparity becomes even more pronounced: the TD computation requires nearly $5.0 \times 10^4$~s, resulting in a speedup exceeding $14$ times. This dramatic slowdown in TD occurs because the relaxation to the periodic limit cycle is physically governed by the viscous diffusion timescale across the annular gap. At high frequencies, the oscillation period is extremely short relative to the global diffusion time, rendering the time-marching solver to resolve many short physical cycles before periodicity is established. 

These results clearly indicate that the HB-UGKS offers significant computational savings over traditional time integration, particularly in high-frequency multiscale regimes where time-accurate tracking of transient periods is prohibitively expensive.

\FloatBarrier
\subsection{Large-Amplitude Thermally Driven Oscillatory Cavity Flow}

To investigate finite-amplitude nonlinear effects, an oscillatory thermally driven flow within a square cavity ($A = 1$) is examined. In contrast to the preceding small-amplitude shear-driven configurations, the bottom, left, and right walls are maintained at a constant reference temperature $T_0$, while the top wall is subjected to a large-amplitude periodic thermal excitation:
\begin{equation}
T_w(t) = T_0 + T_a \cos(\omega t),
\end{equation}
where the temperature amplitude is set to $T_a = 0.3T_0$, and $\omega$ is the oscillation frequency prescribed by the Strouhal number. 

This substantial thermal modulation continuously drives the system far from local thermodynamic equilibrium and excites pronounced higher-order harmonic responses. It thus provides a stringent benchmark for testing both the non-equilibrium kinetic transport and the nonlinear cross-frequency coupling of the HB-UGKS framework. To ensure numerical consistency, the spatial mesh and discrete velocity space remain identical to those employed in the preceding cavity case across all tested Knudsen numbers. Unless otherwise stated, phase-resolved validations are conducted at $St = 2$ across four representative Knudsen regimes ($Kn = 0.01, 0.1, 1.0$, and $10$), followed by a computational efficiency assessment across different frequencies.

\subsubsection{Harmonic independence verification and multiscale spatial structures}

\begin{figure}[htbp]
    \centering
    \includegraphics[width=0.6\linewidth]{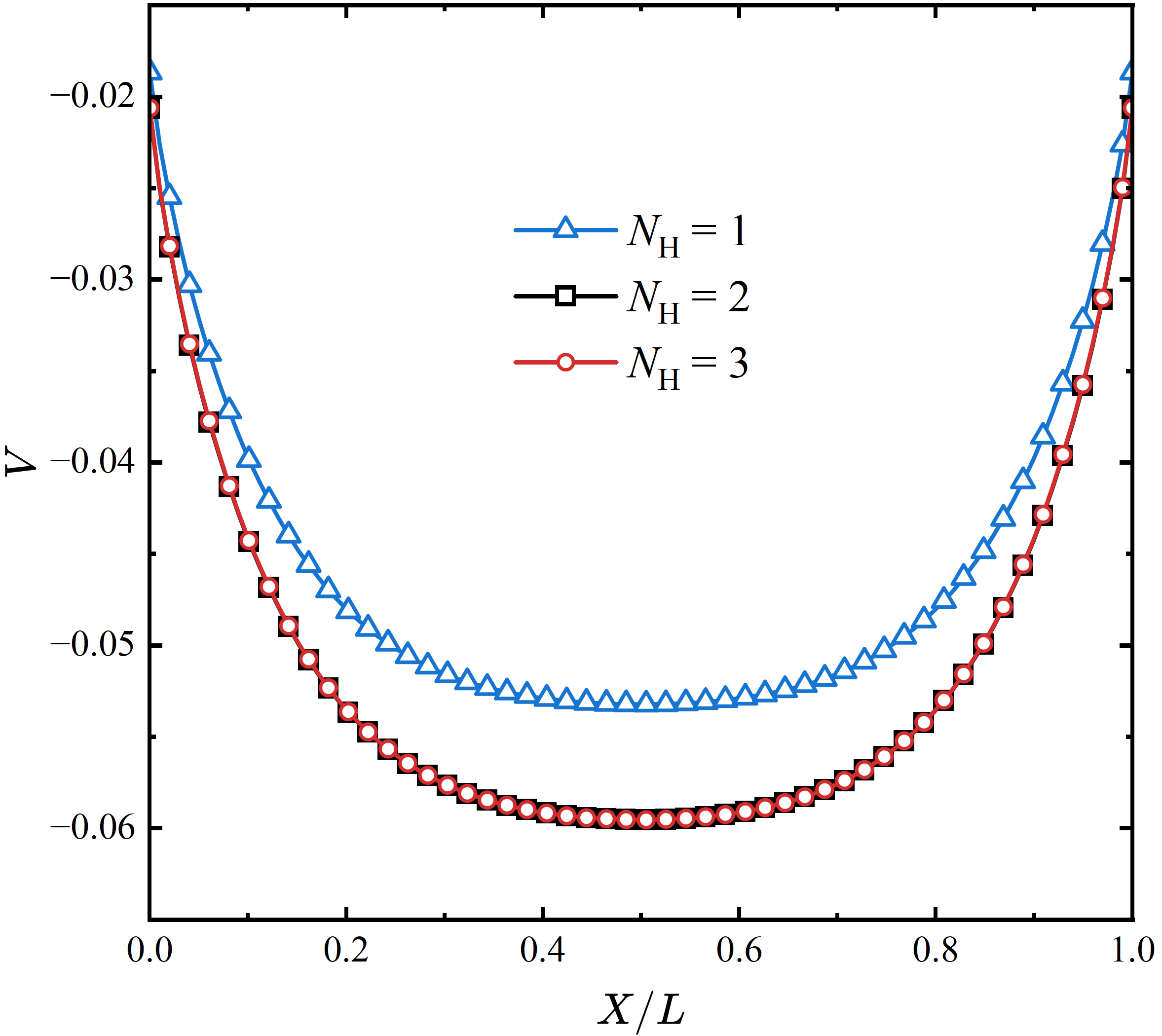}
    \caption{Harmonic truncation study: spatial distribution of vertical velocity $V$ along the horizontal centerline at $t=0.5T$ computed with $N_\text{H} = 1, 2,$ and $3$.}
    \label{fig:harmonic_indep}
\end{figure}

Under finite-amplitude thermal forcing, the nonlinear kinetic and macroscopic terms generate higher-order temporal harmonics that are not present in the imposed single-frequency excitation. Figure~\ref{fig:harmonic_indep} therefore examines the sensitivity of the reconstructed solution to the retained harmonic order. The \(N_{\mathrm H}=1\) result shows a clear deviation, whereas the \(N_{\mathrm H}=2\) and \(N_{\mathrm H}=3\) solutions are nearly indistinguishable. This convergence indicates that the influence of unresolved harmonics beyond the second order on the resolved solution is negligible for the present case. Accordingly, \(N_{\mathrm H}=2\) is adopted for the subsequent thermal simulations.

\begin{figure}[htbp]
    \centering
    \includegraphics[width=0.6\linewidth]{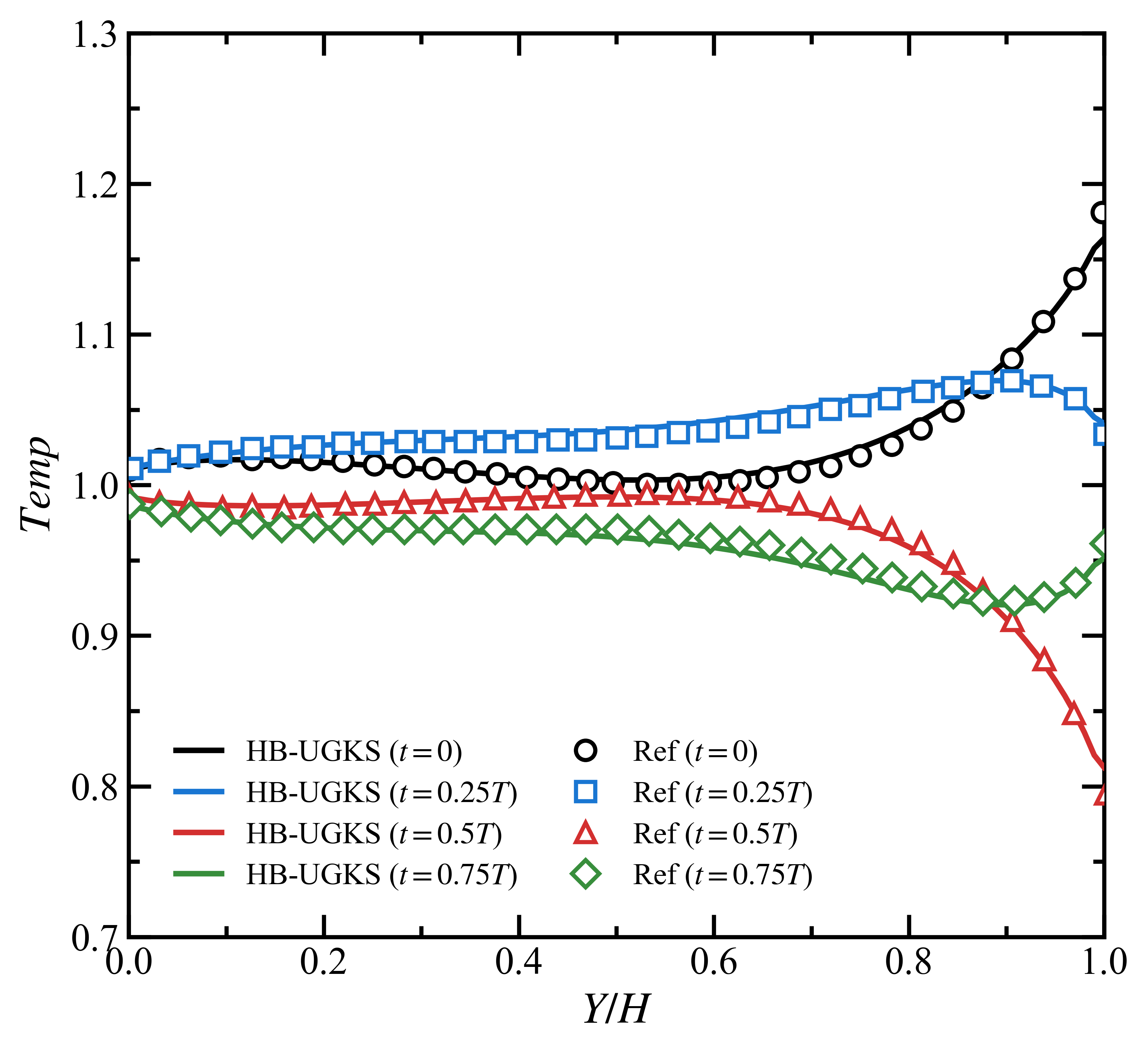}
    \caption{Phase-resolved temperature profiles along the vertical centerline ($x = 0.5L$) over one oscillation period at $Kn=0.1$. Symbols denote reference data from Yang et al.~\cite{Yang2025Hybrid}.}
    \label{fig:T_profiles_Fig4}
\end{figure}

With the harmonic resolution established, the non-equilibrium thermal response in the transitional regime is investigated. Figure~\ref{fig:T_profiles_Fig4} presents the temperature profiles along the vertical centerline at four distinct phases for $Kn = 0.1$.

These profiles highlight key kinetic non-equilibrium characteristics. Most notably, a distinct temperature jump appears adjacent to the oscillating top wall: while the wall temperature reaches extrema of $1.3T_0$ and $0.7T_0$ at $t=0$ and $t=0.5T$, respectively, the adjacent gas temperatures only reach approximately $1.17T_0$ and $0.79T_0$. This macroscopic jump directly reflects non-equilibrium ballistic behavior within the Knudsen layer. As the thermal perturbation propagates downward, it is heavily attenuated by intermolecular collisions, leaving the gas near the bottom wall virtually unperturbed at $T/T_0 \approx 1.0$. Furthermore, finite-rate thermal diffusion induces a noticeable phase lag. For example, at $t=0.25T$, although the wall temperature has returned to $1.0T_0$, the interior gas in the upper cavity ($Y/H \approx 0.6\sim0.9$) remains elevated due to thermal inertia from the preceding heating phase. The HB-UGKS results match the reference kinetic solutions of Yang et al.~\cite{Yang2025Hybrid} with high fidelity, verifying its multiscale accuracy in capturing both kinetic boundary slips and thermal wave propagation.

\begin{figure}[htbp]
    \centering
    \begin{subfigure}{0.48\textwidth}
        \includegraphics[width=\linewidth]{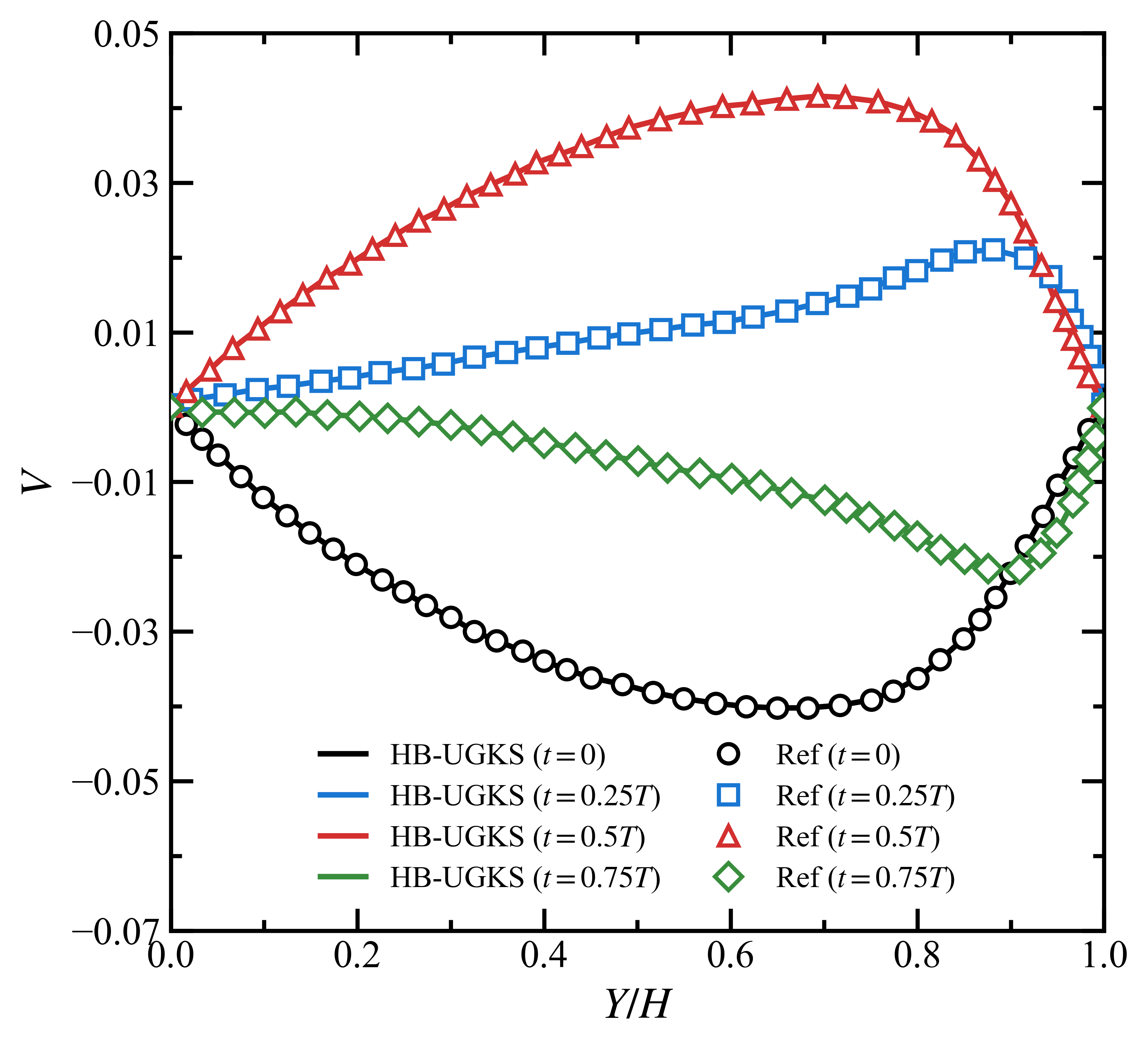}
        \caption{$Kn = 0.01$}
    \end{subfigure}
    \hfill
    \begin{subfigure}{0.48\textwidth}
        \includegraphics[width=\linewidth]{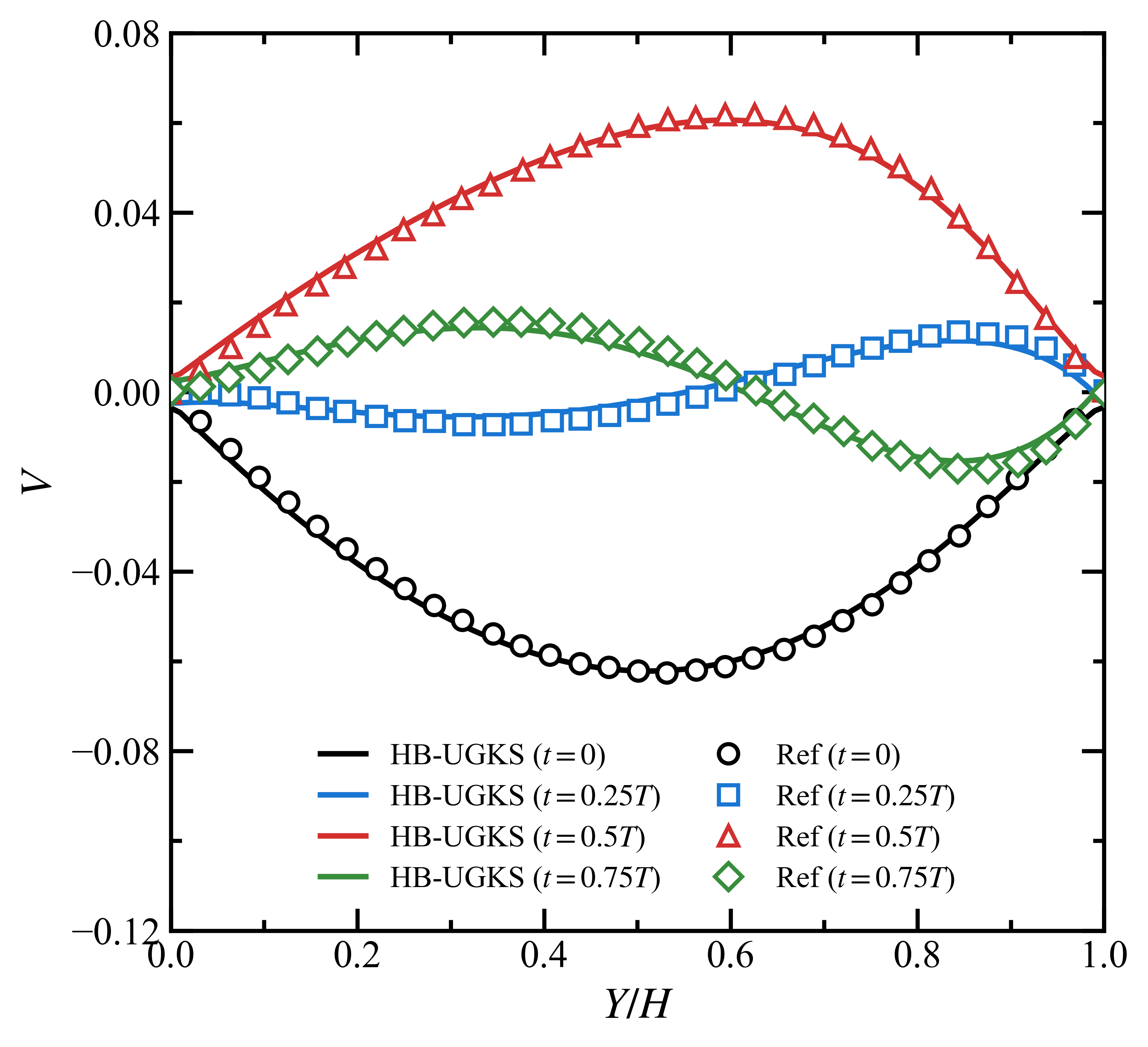}
        \caption{$Kn = 0.1$}
    \end{subfigure}
    \vskip\baselineskip
    \begin{subfigure}{0.48\textwidth}
        \includegraphics[width=\linewidth]{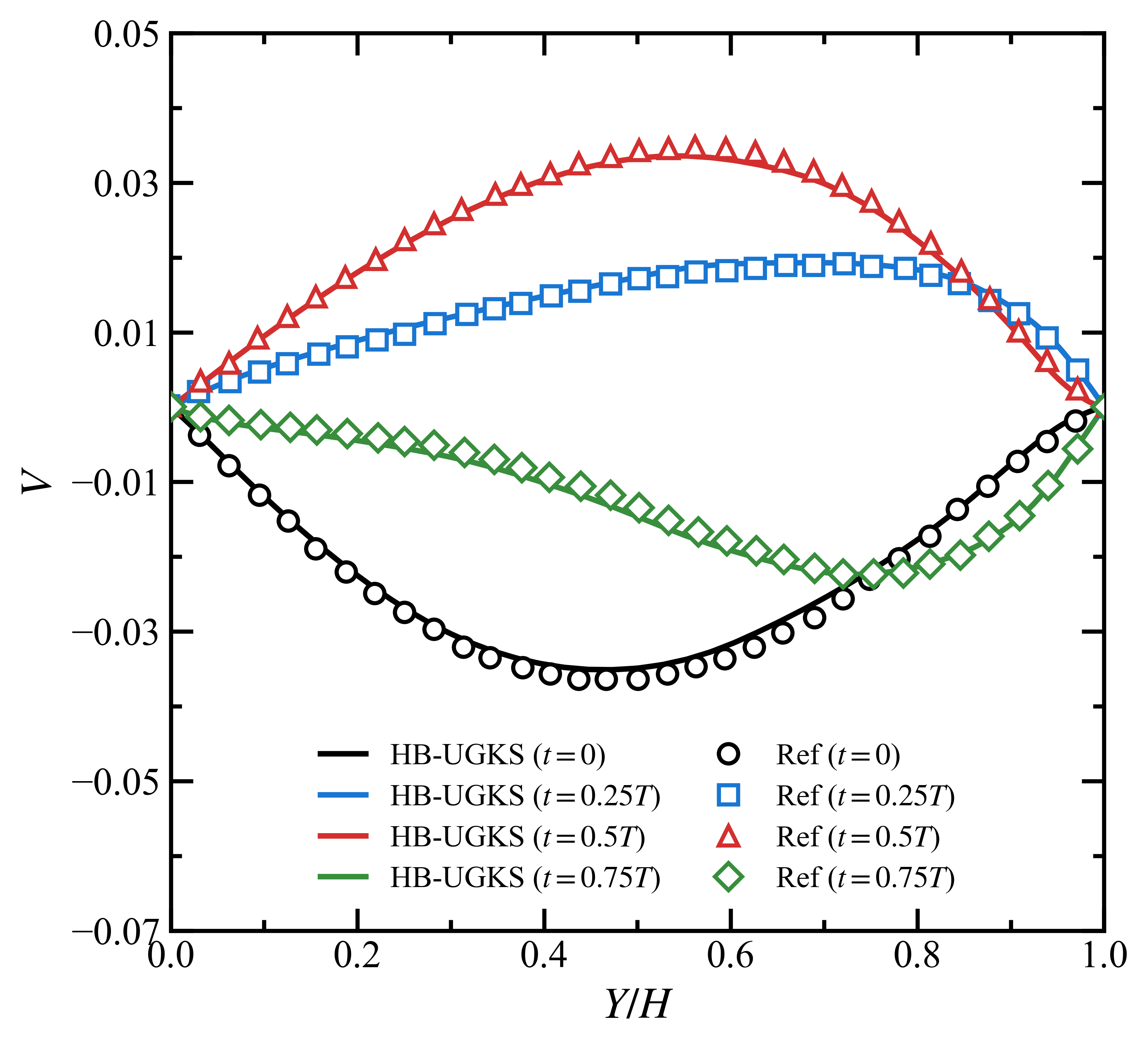}
        \caption{$Kn = 1.0$}
    \end{subfigure}
    \hfill
    \begin{subfigure}{0.48\textwidth}
        \includegraphics[width=\linewidth]{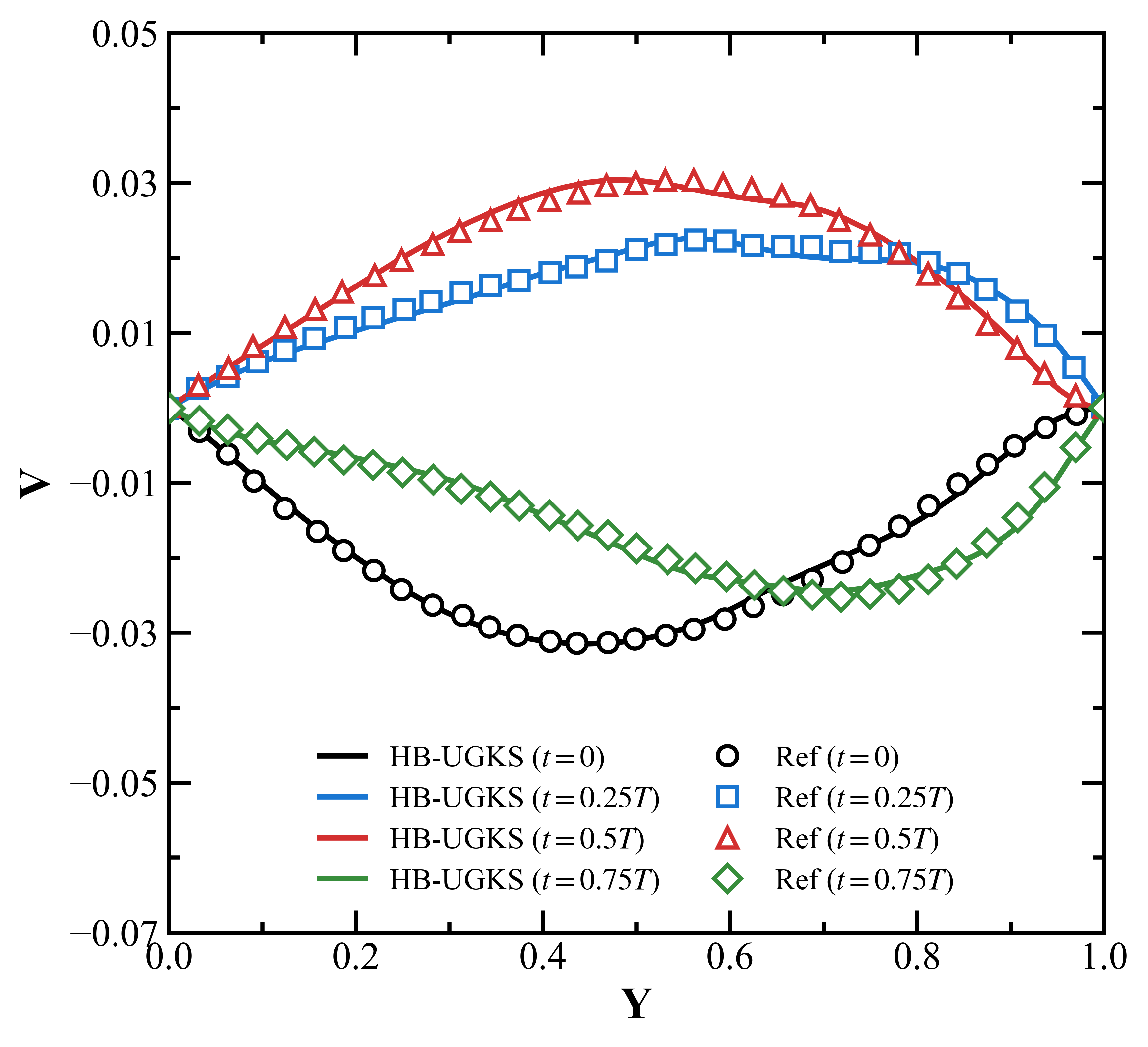}
        \caption{$Kn = 10$}
    \end{subfigure}
    \caption{Vertical velocity ($V$) profiles along the cavity centerline ($x=0.5L$) at different phase angles across various Knudsen numbers. Symbols denote the reference data of Yang et al.~\cite{Yang2025Hybrid}.}
    \label{fig:Velocity_V_Kn}
\end{figure}

The spatial penetration of the oscillatory response across different regimes is illustrated by the vertical velocity profiles in Figure~\ref{fig:Velocity_V_Kn}. At the near-continuum limit ($Kn = 0.01$), the dynamic response is strongly damped by frequent collisions, confining significant macroscopic motion to the upper portion of the cavity. At $Kn = 0.1$, the perturbation penetrates deeper into the core while retaining a collision-dominated, wave-like structure with spatial phase shifts. In contrast, under highly rarefied conditions ($Kn = 1.0$ and $10$), the collision frequency drops drastically; molecules heated by the oscillating boundary travel across the domain with negligible collisions. The collective diffusive wave propagation is thus replaced by ballistic molecular transport. The close agreement between the present HB-UGKS predictions and the reference data across all four Knudsen numbers confirms the robustness of the scheme across continuum, transitional, and free-molecular regimes.

\subsubsection{Phase-Dependent Flow Structures and Nonlinear Temporal Asymmetry}

\begin{figure}[htbp]
    \centering
    \begin{subfigure}{0.48\textwidth}
        \includegraphics[width=\linewidth]{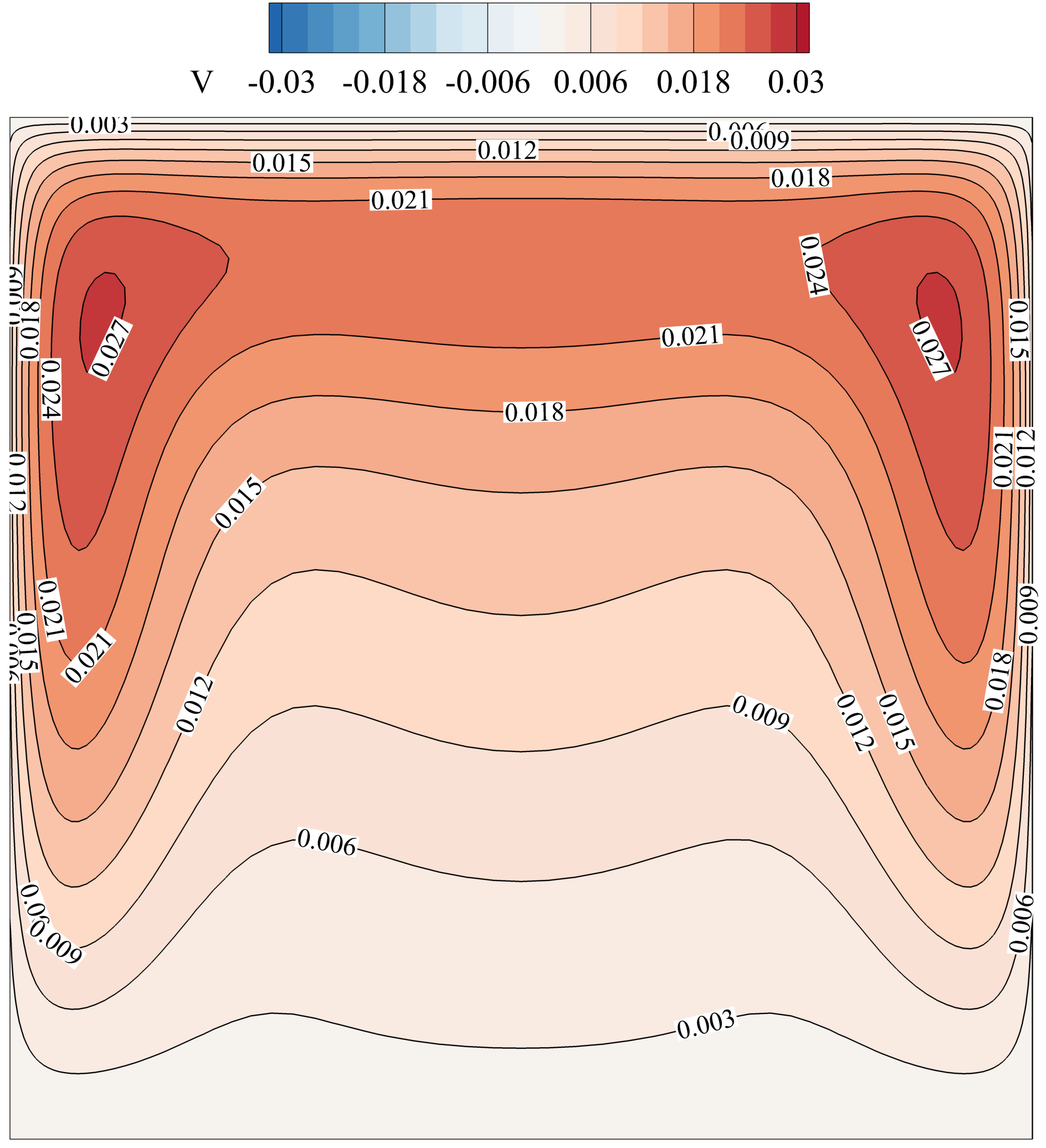}
        \caption{$Kn=0.1,\; t/T=0.25$}
    \end{subfigure}
    \hfill
    \begin{subfigure}{0.48\textwidth}
        \includegraphics[width=\linewidth]{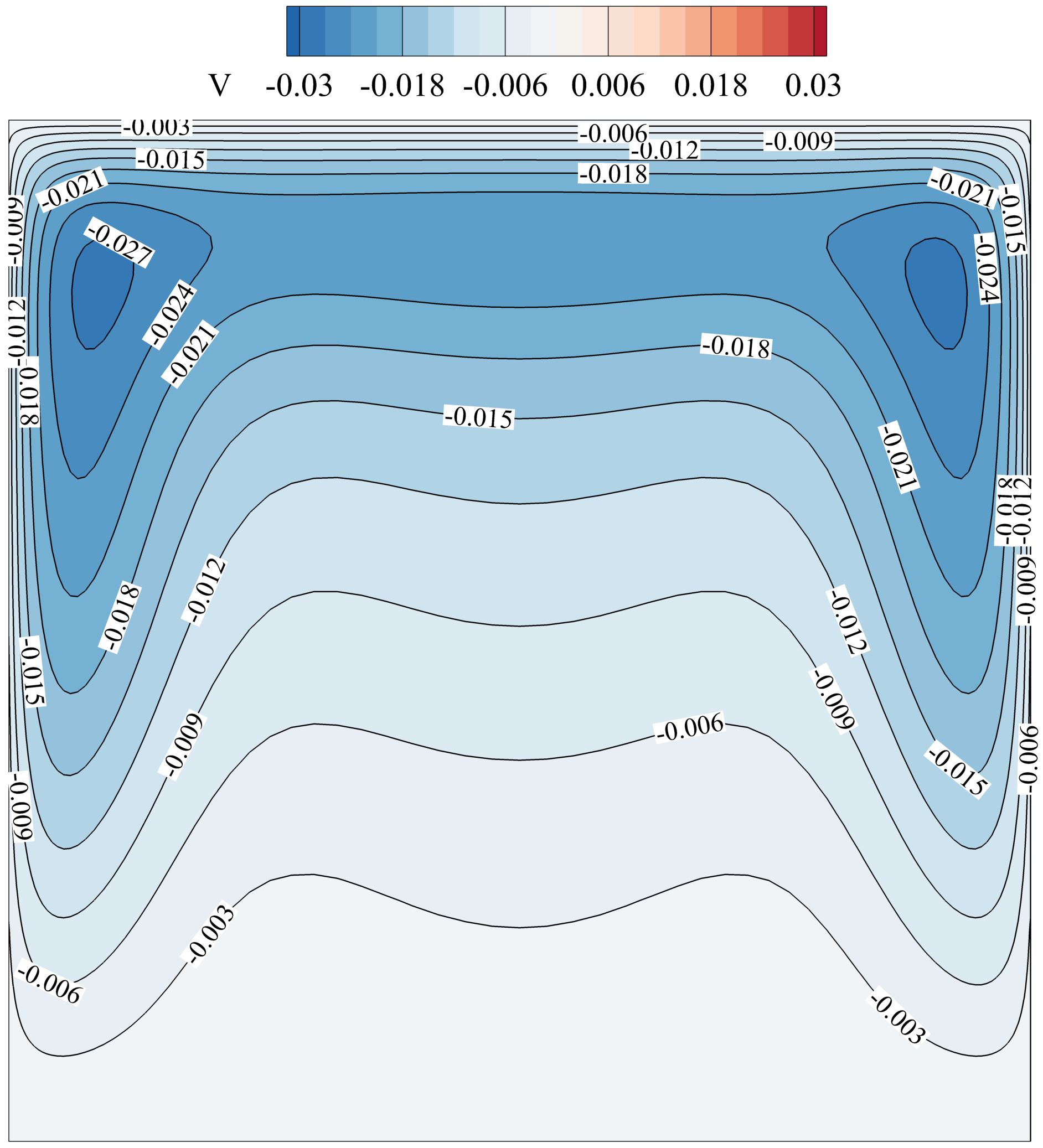}
        \caption{$Kn=0.1,\; t/T=0.75$}
    \end{subfigure}
    \vskip\baselineskip
    \begin{subfigure}{0.48\textwidth}
        \includegraphics[width=\linewidth]{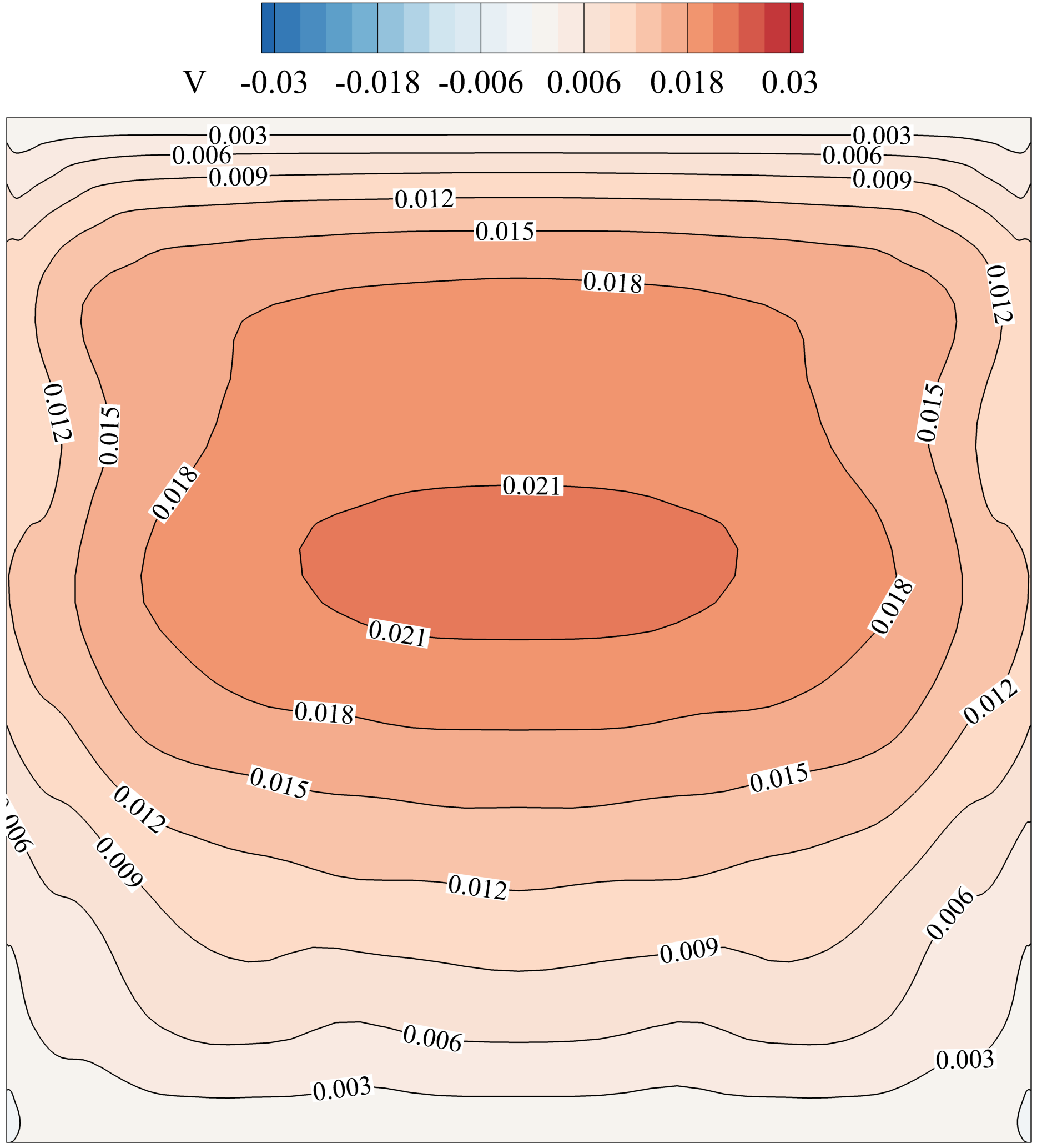}
        \caption{$Kn=10,\; t/T=0.25$}
    \end{subfigure}
    \hfill
    \begin{subfigure}{0.48\textwidth}
        \includegraphics[width=\linewidth]{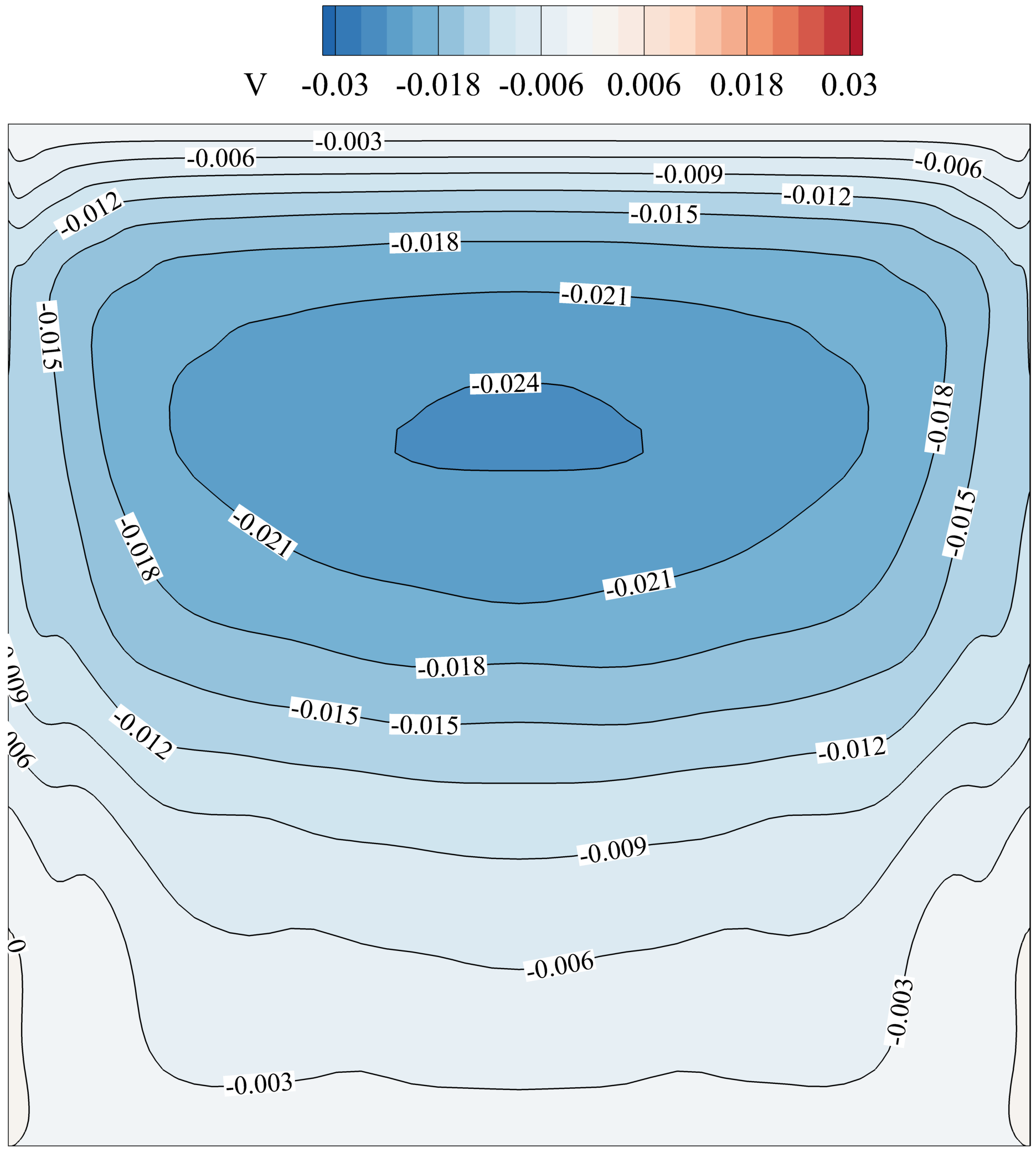}
        \caption{$Kn=10,\; t/T=0.75$}
    \end{subfigure}
    \caption{Phase-resolved contours of the vertical velocity $V$ in the thermally driven cavity at $Kn = 0.1$ and $Kn = 10$.}
    \label{fig:Yang_V_contours}
\end{figure}

The finite-amplitude thermal forcing also alters the internal flow patterns and breaks linear temporal symmetries. Figure~\ref{fig:Yang_V_contours} depicts the phase-resolved vertical velocity contours at two opposing phases: $t/T = 0.25$ (expansion phase) and $t/T = 0.75$ (compression phase). At $Kn = 0.1$, the velocity field remains localized near the driving boundary with roughly comparable magnitudes and opposite flow directions. At $Kn = 10$, however, the velocity disturbances spread across nearly the entire cavity, exhibiting a pronounced breakdown of the linear half-cycle anti-symmetry. This asymmetry originates from the nonlinear dependence of state variables and transport coefficients on temperature, causing the thermal expansion dynamics during heating to differ substantially from the relaxation dynamics during cooling.

\begin{figure}[htbp]
    \centering
    \begin{subfigure}{0.48\textwidth}
        \includegraphics[width=\linewidth]{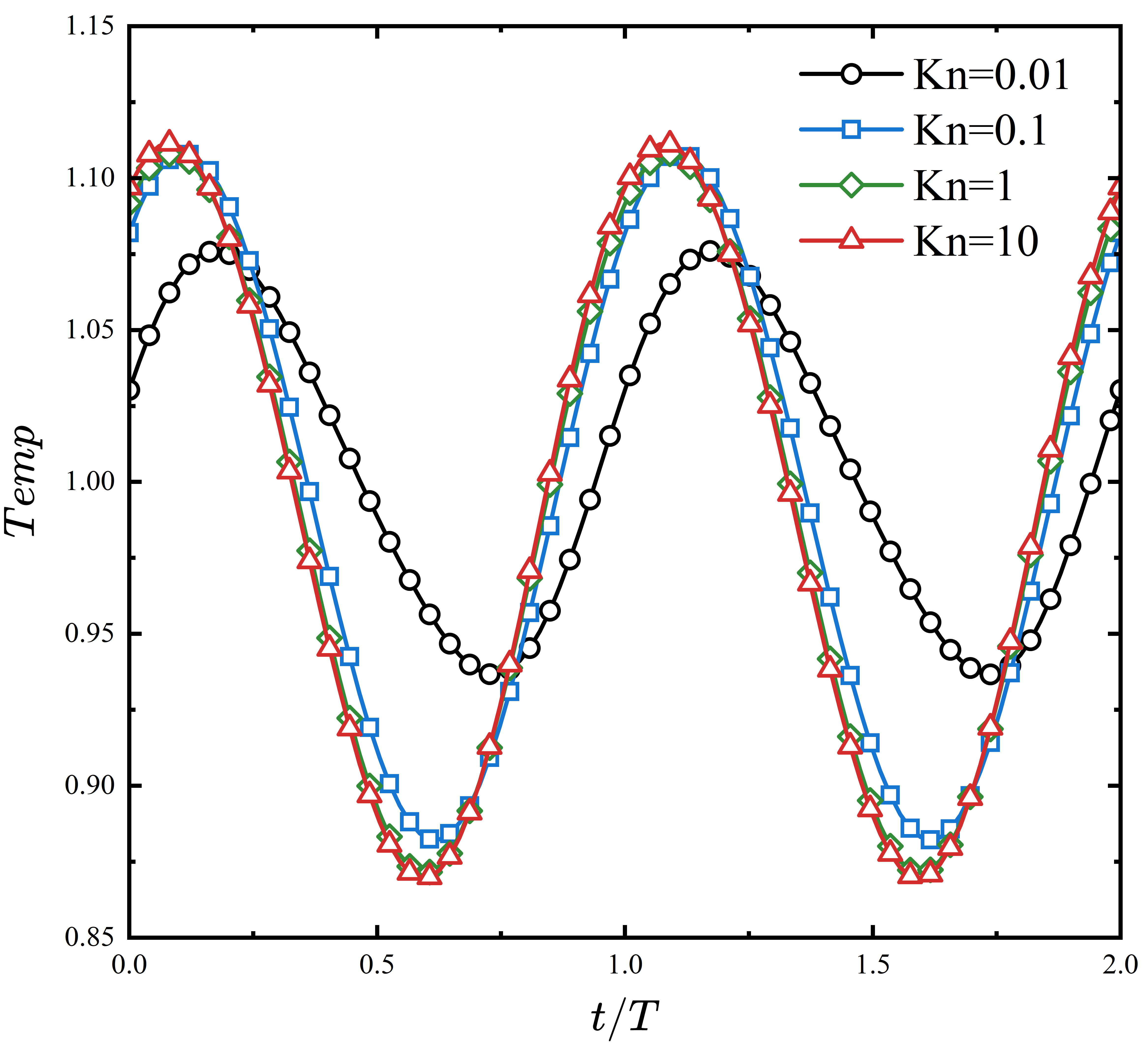}
        \caption{Temperature time history}
    \end{subfigure}
    \hfill
    \begin{subfigure}{0.48\textwidth}
        \includegraphics[width=\linewidth]{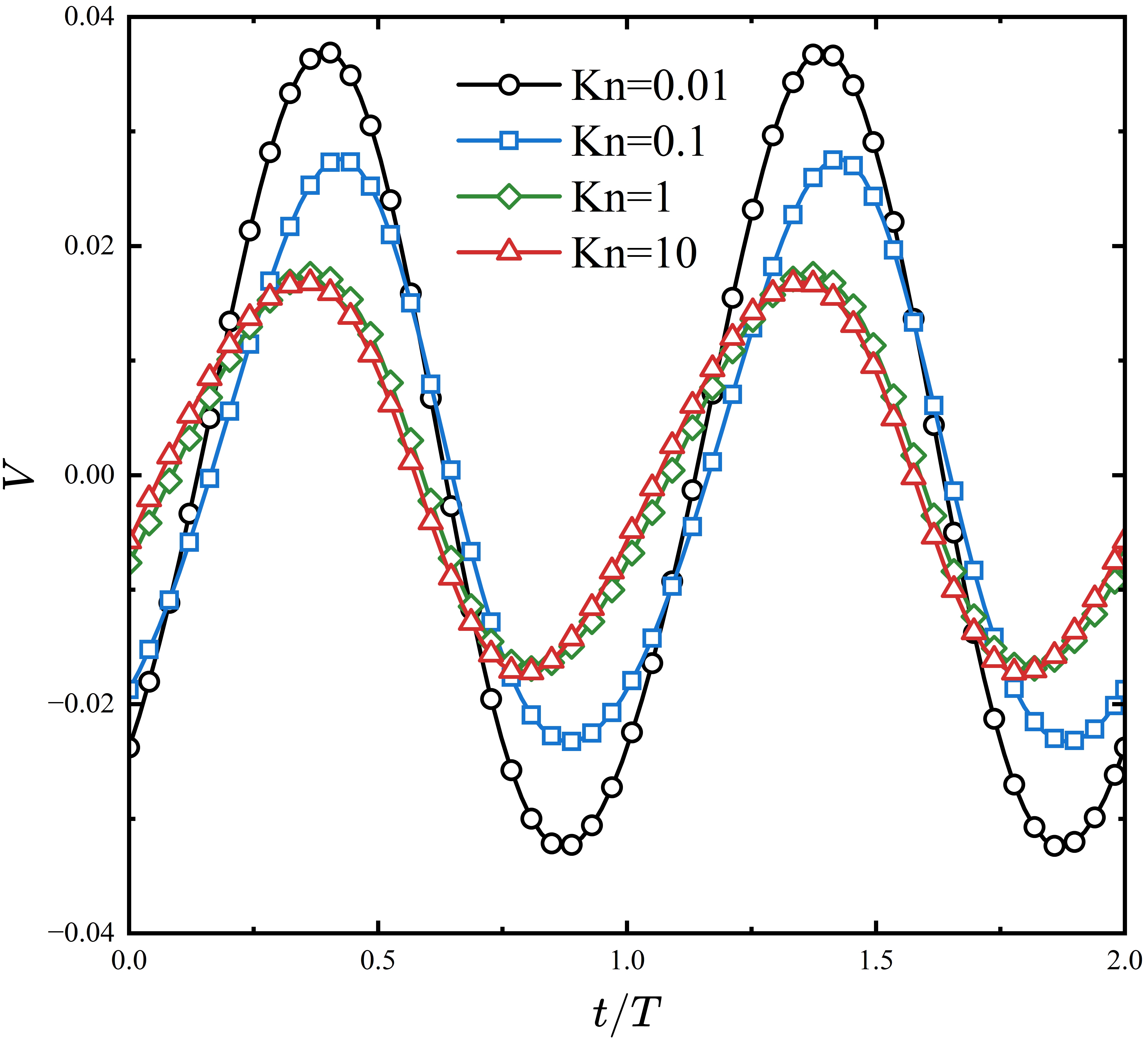}
        \caption{Vertical velocity time history}
    \end{subfigure}
    \caption{Reconstructed temporal evolution of temperature and vertical velocity at probe location ($X=0.5L, Y=0.9H$) over two oscillation periods.}
    \label{fig:recon_time_history}
\end{figure}

The reconstructed temporal histories at the probe point ($X=0.5L, Y=0.9H$) in Figure~\ref{fig:recon_time_history} further quantify this dynamic behavior. The temperature response (Figure~\ref{fig:recon_time_history}a) exhibits an increasing oscillation amplitude with rarefaction while retaining an essentially sinusoidal waveform across all regimes. In contrast, the vertical velocity (Figure~\ref{fig:recon_time_history}b) exhibits an opposite trend in amplitude, along with a distinct non-zero mean shift in the continuum and slip regimes. The positive peak of the velocity fluctuation reaches a noticeably higher absolute value than the negative trough, indicating a net non-zero convective drift induced by nonlinear thermal expansion. As the gas transitions into the free-molecular limit ($Kn=1.0$ and $10$), the overall macroscopic velocity amplitude decreases due to the absence of collective momentum exchange. The HB-UGKS accurately reproduces both the nonlinear mean-shift and the transition in fluctuation amplitudes across all regimes.

\subsubsection{Discussion on Computational Efficiency and Applicability}

The computational efficiency of the HB-UGKS is evaluated against the TD framework within the same underlying numerical framework. Both solvers are marched until the macroscopic density residual satisfies the stopping criterion defined in Eq.~(\ref{eq:benchmark_residuals}). Quantitative comparisons of the reconstructed periodic fields confirm that the HB-UGKS retains high fidelity with respect to the time-domain solutions under the selected harmonic resolutions.

\begin{figure}[htbp]
    \centering
    \includegraphics[width=0.6\linewidth]{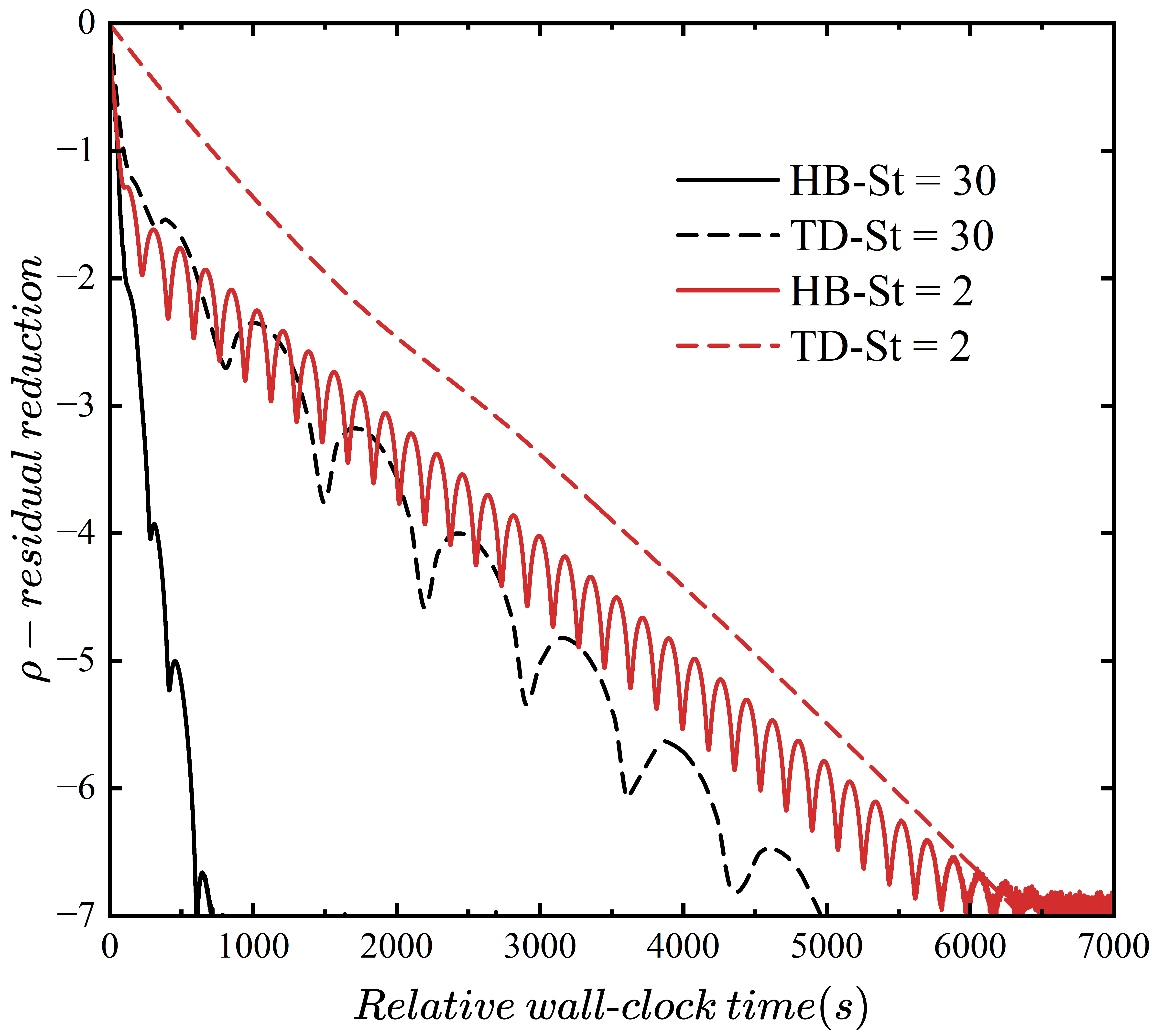}
    \caption{Convergence comparison between the HB-UGKS and the explicit time-domain UGKS for the thermally driven cavity under the benchmark criterion of Eq.~(\ref{eq:benchmark_residuals}). The ordinate is the shifted density-residual reduction $\Delta r_\rho$, and the abscissa is the shifted wall-clock time $\Delta t_{\mathrm{wall}}$.}
    \label{fig:yang_speedup}
\end{figure}

Figure~\ref{fig:yang_speedup} compares the wall-clock convergence histories of both solvers at two representative frequencies.

At the low-frequency condition ($St=2$), the HB solver achieves a modest speedup of approximately $1.06\times$, compared with approximately $1.4\times$ for the linear shear-driven cavity at the same $St$. The smaller gain is consistent with the additional computational overhead associated with the higher harmonic resolution. In the thermally driven case, retaining $N_{\mathrm H}=2$ requires 5 coupled temporal
collocation states, increasing the storage requirements and the cost of flux evaluations and spectral coupling at each pseudo-time step. Meanwhile, the explicit TD solver remains competitive at low frequencies, where relatively few physical cycles are needed to eliminate startup transients.

At the higher frequency ($St=30$), however, the computational advantage of the frequency-domain formulation becomes pronounced, achieving an acceleration factor of approximately $8.33\times$. Because the intrinsic physical relaxation time of the cavity flow is dictated by molecular diffusion and collision timescales rather than the excitation period, the TD solver must resolve many short physical cycles to wash out initial transients. In contrast, the HB method solves directly for the periodic limit cycle in pseudo-time, rendering its iteration count largely independent of the physical period length. This demonstrates the efficiency and robustness of the present framework for high-frequency non-equilibrium periodic flows.

A few remarks are in order regarding the efficiency comparison and the practical applicability of the HB-UGKS framework. In this study, both the TD and HB solvers are evaluated within an identical algorithmic discretization to strictly isolate the acceleration effect provided by the harmonic balance formulation. While an explicit scheme is adopted as the baseline time integration here, the HB framework reformulates the unsteady problem into a coupled quasi-steady system. Consequently, standard steady acceleration techniques—such as implicit macroscopic-microscopic iterations and geometric multigrid—can be inherited directly to achieve further convergence acceleration.

On the other hand, a realistic trade-off lies in the computational memory and algebraic coupling. For kinetic methods, the phase space already carries a substantial dimension due to the velocity-space discretization. Under strong flow nonlinearities where a large harmonic truncation $N_\text{H}$ is required, storing and coupling distribution functions across $2N_\text{H}+1$ temporal collocation points noticeably increases the memory footprint and the arithmetic cost per pseudo-time step. Therefore, the present HB-UGKS is particularly advantageous for weakly to moderately nonlinear multiscale problems, such as high-frequency MEMS/NEMS resonators and oscillatory boundary layers. In these applications, a low harmonic truncation ($N_\text{H} = 1\sim 2$) is sufficient to capture the dominant physical dynamics, while the decoupled structure of each temporal collocation point facilitates straightforward parallel implementation and substantial wall-clock savings.

\FloatBarrier
\section{Conclusions}
\label{sec:conclusions}

In this work, a time-domain harmonic balance unified gas-kinetic scheme (HB-UGKS) is developed for the efficient simulation of temporally periodic flows across all Knudsen regimes. The framework employs a spectral operator to treat the time derivatives in both the microscopic kinetic and macroscopic governing equations, while inheriting the multiscale flux evaluation of the UGKS to capture non-equilibrium transport across disparate flow regimes. By transforming the original unsteady governing equations into a block-coupled, quasi-steady system, the periodic limit cycle is directly resolved via explicit pseudo-time marching combined with local time-stepping.

The numerical performance and physical fidelity of the framework are demonstrated through three complementary benchmarks that systematically separate geometric complexity from finite-amplitude nonlinearity. First, the small-amplitude shear-driven cavity validates the baseline algorithm, demonstrating harmonic truncation convergence, aspect-ratio anti-resonance scaling, and proper limiting behavior from the continuum to the collisionless regime. Next, the eccentric annulus problem confirms the capability of the scheme on curved boundaries, accurately capturing the frequency-dependent dynamic shear traction and the non-equilibrium load overshoot over a wide frequency spectrum up to $S = 30$. Finally, the large-amplitude thermally driven cavity addresses finite-amplitude nonlinearities, showing that retaining second-order harmonics ($N_\text{H} = 2$) is both necessary and sufficient to capture dynamic half-cycle asymmetry and nonlinear acoustic streaming effects.

Quantitative efficiency comparisons against explicit time-domain simulations show that the speedup provided by the HB-UGKS depends strongly on both the excitation frequency and the degree of flow
nonlinearity. In the present test cases, the low-frequency speedup ranges from about $1.06\times$ in the nonlinear thermally driven cavity to about $1.4\times$ in the linear shear-driven cavity, while
the eccentric-annulus case already reaches about $6\times$ at $S=1$. At higher frequencies, the advantage becomes much more pronounced, with representative speedups of about $8.3\times$ in the
thermally driven cavity, $9.2\times$ in the shear-driven cavity, and more than $14\times$ in the eccentric annulus. These results confirm that the HB-UGKS is particularly attractive for high-frequency
periodic multiscale flows, where conventional time-domain marching must resolve many short cycles before reaching the periodic limit state.

Overall, the HB-UGKS provides a robust and efficient approach for simulating periodic multiscale flows by entirely bypassing physical startup transients. This efficiency makes the method particularly attractive for high-frequency, moderately nonlinear applications such as micro-electromechanical resonators. Future efforts will explore implicit macro-micro iterations to further accelerate pseudo-time convergence, along with extensions to multi-frequency and three-dimensional configurations.

\section*{Acknowledgments}

The authors would like to express their sincere gratitude to Prof. Dingxi WANG from the School of Power and Energy at Northwestern Polytechnical University for his insightful discussions and valuable suggestions regarding the harmonic balance method. We are also grateful to Mr. Junzhe CAO, Dr. Xiaojian YANG, Dr. Yue ZHANG and Dr. Zhigang PU for their helpful assistance with the technical details during the implementation of this work.
The current research is supported by National Key R\&D Program of China (Grant No. 2022YFA1004500), National Natural Science Foundation of China (92371107), and Research Grants Council of Hong Kong (16208324).

\bibliographystyle{elsarticle-num}
\bibliography{elsarticle-num.bib}
\end{document}